\documentclass{aa}

\usepackage{graphicx}
\usepackage{epsfig}
\usepackage[outdir=./Figures]{epstopdf}
\usepackage{textcomp}
\usepackage{gensymb}
\usepackage{xcolor}
\usepackage{txfonts}
\bibpunct{(}{)}{;}{a}{}{,} 

\usepackage{acro}

\acsetup{
  single         = false,
  sort           = true,
  group-citation = true,
  group-cite-cmd = \citealp,
  group-cite-connect = {; },
}

\DeclareAcronym{EVE}{
  short = EVE,
  long = \textit{Extreme Ultraviolet Variability Experiment},
  cite = {Woods:2012}}

\DeclareAcronym{AIA}{
  short = AIA,
  long = \textit{Atmospheric Imaging Assembly},
  cite = {Lemen:2012}}

\DeclareAcronym{HMI}{
  short = HMI,
  long = \textit{Helioseismic Magnetic Imager},
  cite = {Scherrer:2012}}
  
\DeclareAcronym{SDO}{
  short = SDO,
  long = \textit{Solar Dynamics Observatory},
  cite = {Pesnell:2012}}

\DeclareAcronym{STEREO}{
  short = STEREO,
  long = \textit{Solar Terrestrial Relations Observatory},
  cite = {Kaiser:2008}}

\DeclareAcronym{SECCHI}{
  short = SECCHI,
  long = \textit{Sun Earth Connection Coronal and Heliospheric Investigation},
  cite = {Howard:2008}}
  
\DeclareAcronym{SCIP}{
  short = SCIP,
  long = Sun Centered Imaging Package}
  
\DeclareAcronym{MGN}{
  short = MGN,
  long = multi-scale Gaussian normalisation,
  cite = {Morgan:2014}}

\DeclareAcronym{EUVI}{
  short = EUVI,
  long = EUVI}

\DeclareAcronym{CME}{
  short = CME,
  short-plural-form = CMEs,
  long = coronal mass ejection,
  long-plural-form = coronal mass ejections}
  
\DeclareAcronym{PIL}{
  short = PIL,
  short-plural-form = PILs,
  long = polarity inversion line,
  long-plural-form = polarity inversion lines}
  
\DeclareAcronym{LOS}{
  short = LOS,
  short-plural-form = LOS,
  long = line of sight,
  long-plural-form = lines of sight}
  
\DeclareAcronym{FOV}{
  short = FOV,
  short-plural-form = FOV,
  long = field of view,
  long-plural-form = fields of view}
  
\DeclareAcronym{EUV}{
  short = EUV,
  short-plural-form = EUV,
  long = extreme ultraviolet,
  long-plural-form = extreme ultraviolet}
  
\DeclareAcronym{GONG}{
  short = GONG,
  long = \textit{Global Oscillations Network Group}}
  
\DeclareAcronym{DST}{
  short = DST,
  long = \textit{Dunn Solar Telescope}}

\DeclareAcronym{IBIS}{
  short = IBIS,
  long = \textit{Interferometric Bidimensional Spectropolarimeter},
  cite = {Cavallini:2006,Reardon:2008}}

\DeclareAcronym{FIRS}{
  short = FIRS,
  long = \textit{Facility Infrared Spectropolarimeter},
  cite = {Jaeggli:2011}}
  
\DeclareAcronym{EIT}{
  short = EIT,
  long = \textit{Extreme Ultraviolet Imaging Telescope},
  cite = {Delaboudiniere:1995}}
 
\DeclareAcronym{ROSA}{
  short = ROSA,
  long = \textit{Rapid Oscillations in the Solar Atmosphere},
  cite = {Jess:2010}}
  
\DeclareAcronym{SPINOR}{
  short = SPINOR,
  long = \textit{Spectro-Polarimeter for Infrared and Optical Regions},
  cite = {Socas:2006}}  
 
\DeclareAcronym{SOHO}{
  short = SOHO,
  long = \textit{Solar and Heliospheric Observatory},
  cite = {Domingo:1995}}
 
\DeclareAcronym{MDI}{
  short = MDI,
  long = \textit{Michelson Doppler Imager},
  cite = {Scherrer:1995}}

\DeclareAcronym{NMSU}{
  short = NMSU,
  long = New Mexico State University}

\DeclareAcronym{NSO}{
  short = NSO,
  long = \textit{National Solar Observatory}}
  
\DeclareAcronym{TAC}{
  short = TAC,
  long = time allocation committee}
  
\DeclareAcronym{DEM}{
  short = DEM,
  short-plural-form = DEMs,
  long = Differential Emission Measure,
  long-plural-form = Differential Emission Measures}
  
\DeclareAcronym{PI}{
  short = PI,
  short-plural-form = PI,
  long = Principle Investigator,
  long-plural-form = Principle Investigators}  
  
\DeclareAcronym{IDL}{
  short = IDL,
  long = Interactive Data Language}
  
\DeclareAcronym{bb}{
  short = bb,
  long = broadband}  
  
\DeclareAcronym{nb}{
  short = nb,
  long = narrowband}    
  
\DeclareAcronym{IPM}{
  short = IPM,
  long = interplanetary medium}  
  
\DeclareAcronym{AO}{
  short = AO,
  long = adaptive optics,
  cite = {Rimmele:2004}} 
  
\DeclareAcronym{IR}{
  short = IR,
  long = infrared}  

\DeclareAcronym{halpha}{
  short = H-$\alpha$,
  long = Hydrogen-$\alpha$}

\DeclareAcronym{FPI}{
  short = FPI,
  long = Fabry-P\'erot}
  
\DeclareAcronym{DWDM}{
  short = DWDM,
  long = dense wavelength division multiplexing}  

\DeclareAcronym{CCD}{
  short = CCD,
  long = charge-coupled device}   
  
\DeclareAcronym{HI}{
  short = HI,
  long = Heliospheric Investigation}   
  
\DeclareAcronym{ROI}{
  short = ROI,
  long = region-of-interest}    
  
\DeclareAcronym{POV}{
  short = POV,
  long = point-of-view}  
  
\DeclareAcronym{MHS}{
  short = MHS,
  long = magnetohydrostatic}
  
\DeclareAcronym{MHD}{
  short = MHD,
  long = magnetohydrodynamic}  
  
\DeclareAcronym{DOT}{
  short = DOT,
  long = \textit{Dutch Open Telescope},
  cite = {Rutten:1997}}  
  
\DeclareAcronym{BBSO}{
  short = BBSO,
  long = \textit{Big Bear Solar Observatory}}
  
\DeclareAcronym{NST}{
  short = NST,
  long = \textit{New Solar Telescope},
  cite = {Goode:2012}}

\DeclareAcronym{SOT}{
  short = SOT,
  long = \textit{Solar Optical Telescope},
  cite = {Tsuneta:2008}}
  
\DeclareAcronym{hinode}{
  short = Hinode,
  long = Hinode,
  cite = {Kosugi:2007}}
  
\DeclareAcronym{GST}{
  short = GST,
  long = \textit{Goode Solar Telescope}}  
  
\DeclareAcronym{SST}{
  short = SST,
  long = \textit{Swedish 1-m Solar Telescope},
  cite = {Scharmer:2002}}  
  
\DeclareAcronym{TRACE}{
  short = TRACE,
  long = \textit{Transition Region and Coronal Explorer},
  cite = {Handy:1999}}  
  
\DeclareAcronym{PCTR}{
  short = PCTR,
  long = \textit{prominence-corona-transition-region}}
  
\DeclareAcronym{DKIST}{
  short = DKIST,
  long = Daniel K. Inouye Solar Telescope}
  
\DeclareAcronym{RTI}{
  short = RTI,
  long = Rayleigh-Taylor instability}
  
\DeclareAcronym{SSW}{
  short = SSW,
  long = SolarSoftWare,
  cite = {Freeland:1998}}    
  
\DeclareAcronym{LTE}{
  short = LTE,
  long = local thermodynamic equilibrium}
  
\DeclareAcronym{NLTE}{
  short = NLTE,
  long = non-local thermodynamic equilibrium}
  
\DeclareAcronym{FTS}{
  short = FTS,
  long = Fourier Transform Spectrometer,
  cite = {Kurucz:1984}}
  
\DeclareAcronym{RTE}{
  short = RTE,
  long = radiative transfer equation}
  
\DeclareAcronym{CRD}{
  short = CRD,
  long = complete redistribution}
  
\DeclareAcronym{PRD}{
  short = PRD,
  long = partial redistribution}
  
\DeclareAcronym{BCM}{
	short = BCM,
	long = Beckers' cloud model,
	cite={Beckers:1964}}
	
\DeclareAcronym{HSRA}{
	short = HSRA,
	long = Harvard Smithsonian Reference Atmosphere,
	cite={Gingerich:1971}}
	
\DeclareAcronym{NICOLE}{
	short = NICOLE,
	long = Non-LTE Inversion COde using the Lorien Engine,
	cite={Socasnavarro:2015}}
	
\DeclareAcronym{SIR}{
	short = SIR,
	long = Stokes Inversion based on Response functions,
	cite={Ruizcobo:1992}}
	
\DeclareAcronym{HAZEL}{
	short = HAZEL,
	long = Hanle and Zeeman Light,
	cite={AsensioRamos:2008}}
	
\DeclareAcronym{BPSS}{
	short = BPSS,
	long = bald patch separatrix surface,
	cite={Bungey:1996}}
	
\DeclareAcronym{EIS}{
	short = EIS,
	long = \textit{EUV Imaging Spectrometer},
	cite={Culhane:2007}}

\DeclareAcronym{FWHM}{
  short = FWHM,
  long = full width at half maximum}

\DeclareAcronym{AMRVAC}{
	short = MPI-AMRVAC,
	long = \textit{Adaptive Mesh Refinement Versatile Advection Code},
	cite = {Keppens:2012,Porth:2014,Xia:2018,Keppens:2020}}

\DeclareAcronym{AMR}{
	short = AMR,
	long = adaptive mesh refinement}

\DeclareAcronym{CCI}{
	short = CCI,
	long = Convective Continuum Instability}

\DeclareAcronym{BV}{
	short = BV,
	long = Brunt-V\"ais\"al\"a}

\DeclareAcronym{TVDLF}{
	short = TVDLF,
	long = Total Variation Diminishing Lax-Friedrich}

\DeclareAcronym{TI}{
	short = TI,
	long = Thermal Instability}

\DeclareAcronym{TNE}{
	short = TNE,
	long = thermal non-equilibrium}

\makeatletter
\newcommand{\mypm}{\mathbin{\mathpalette\@mypm\relax}}
\newcommand{\@mypm}[2]{\ooalign{%
  \raisebox{.1\height}{$#1+$}\cr
  \smash{\raisebox{-.6\height}{$#1-$}}\cr}}
\makeatother
 
\usepackage[flushleft]{threeparttable}
\usepackage[referable]{threeparttablex}
\usepackage{tabularx}
\usepackage{amsmath}
\usepackage{amsfonts}
\usepackage{amssymb}
\usepackage{mathrsfs}
\usepackage[utf8]{inputenc}
\DeclareUnicodeCharacter{2212}{-}

\usepackage[pdftex,breaklinks, colorlinks, citecolor=blue, linkcolor=blue]{hyperref}

\usepackage{breakcites}

\usepackage{physics}

\usepackage{booktabs}

\usepackage{multirow}

\definecolor{ao(english)}{rgb}{0.0, 0.5, 0.0} 

\newcommand{\BE}{\begin{equation}}
\newcommand{\EE}{\end{equation}}
\newcommand{\BA}{\begin{eqnarray}}
\newcommand{\EA}{\end{eqnarray}}

\begin{document}

	\title{Prominence formation by levitation-condensation at extreme resolutions}

   \author{J. M. Jenkins
          \inst{1}
          \and
          R. Keppens
          \inst{1}
          }

   \institute{Centre for mathematical Plasma-Astrophysics, Celestijnenlaan 200B, 3001 Leuven, KU Leuven, Belgium\\
            \email{jack.jenkins@kuleuven.be}
        }

   \date{Received 2020; accepted 2020}

 
  \abstract
   { Prominences in the solar atmosphere represent an intriguing and delicate balance of forces and thermodynamics in an evolving magnetic topology. How this relatively cool material comes to reside at coronal heights, and what drives its evolution prior to, during, and after its appearance remains an area full of open questions.}
   { We here set forth to identify the physical processes driving the formation and evolution of prominence condensations within 2.5D magnetic flux ropes. We deliberately focus on the levitation-condensation scenario, where a coronal flux rope forms and eventually demonstrates in-situ condensations, revisiting it at extreme resolutions down to order 6 km in scale.}
   { We perform grid-adaptive numerical simulations in a 2.5D translationally invariant setup, where we can study the distribution of all metrics involved in advanced magnetohydrodynamic stability theory for nested flux rope equilibria. We quantify in particular \acf{CCI}, \acf{TI}, baroclinicity, and mass-slipping metrics within a series of numerical simulations of prominences formed via levitation-condensation.}
   {Overall, we find that the formation and evolution of prominence condensations happens in a clearly defined sequence regardless of resolution or background field strength between 3 and 10 Gauss. The \ac{CCI} governs the slow evolution of plasma prior to the formation of distinct condensations found to be driven by the \ac{TI}. Evolution of the condensations towards the topological dips of the magnetic flux rope is a consequence of these condensations forming initially outside of pressure balance with their surroundings. From the baroclinicity distributions, smaller-scale rotational motions are inferred within forming and evolving condensations. Upon the complete condensation of a prominence `monolith', the slow descent of this plasma towards lower heights appears consistent with the mass-slippage mechanism driven by episodes of both local current diffusion and magnetic reconnection.}
   {}


   \maketitle
%
\section{Introduction}\label{s:intro}
The mechanisms suggested to be responsible for the formation of solar prominences within the million degree solar corona are historically split into three main processes, levitation, injection, and condensation \citep[cf.][]{Mackay:2010}. The `levitation' mechanism thereby describes the lifting of already-cool, but very heavy material \textit{i.e.}, plasma-$\beta \gtrapprox 1$, from the solar chromosphere up into coronal heights. In that process, the  magnetic dips are already present and provide the upward balance against gravity \citep[see \textit{e.g.},][]{Kippenhahn:1957,vanBallegooijen:1989,Zhao:2017,Zhao:2019,Zhao:2020}. The `injection' mechanism usually invokes some magnetic reconnection within the chromospheric volume so as to propel material into coronal heights. Sustained capture and suspension of this propelled plasma is then only achieved if the correct topology \textit{already} exists within the corona and is located along those field lines with propelled material, otherwise this material would simply fall back to the chromosphere \citep[][]{Wang:1999b, Chae:2001}. The final method, `condensation', takes advantage of the optically-thin emission processes occuring within the coronal volume \citep[as eluded to by][]{Kiepenheuer:1953}. Herein, coronal plasma cools into condensations that collect within, or themselves create, the topological dips required for their longer-term stability \citep[][]{Antiochos:1991}. In recent years, modern variants or mixtures of these mechanisms came into focus, with especially the `evaporation-condensation' process invoking chromospheric plasma evaporations due to localized, sustained footpoint heating, followed by in-situ condensations in arcades and flux ropes \citep[\textit{e.g.},][]{Xia:2014, Keppens:2014, Xia:2016}. In this work, we revisit the coronal `levitation-condensation' mechanism proposed in \citet{Kaneko:2015}, where a coronal-only volume is considered, where a flux rope forms, coronal plasma gets lifted, and finally condenses in-situ due to thermal instability (TI).

The condensation and TI mechanism has been studied extensively over the last half century \citep[][]{Parker:1953, Field:1965}. However, for prominences it was quickly realised that this in-situ condensation process alone could not provide enough plasma to the prominence to explain their characteristic masses, since the associated coronal volume simply does not contain sufficient mass \citep[\textit{e.g.},][]{Saito:1973}. Subsequently, authors explored pairing the condensation mechanism with localised heating at the footpoints of the associated magnetic fields. Such approaches are also commonly invoked in the \ac{TNE}-cycle models to explain the mass-supply to the coronal rain phenomenon \citep[cf.][]{Klimchuk:2019b,Antolin:2020}. These \ac{TNE} models have thus far been explored parametrically in 1D hydrodynamic settings, and require a magnetic connectivity to the lower chromosphere along with parametrized heating near transition region heights. The resulting additional `evaporation' introduces a method of providing more material to the coronal volume without the requirement for any reconnection, as is needed in the injection model \citep[][]{Antiochos:1999b, Klimchuk:2019b}. The continuous supply of material to the coronal volume in this way overcomes the limitations of the condensation model. Indeed, this `evaporation/condensation' mechanism has been adopted as the basis for many recent numerical simulations of coronal rain \citep[\textit{e.g.},][]{Fang:2013, Moschou:2015, Xia:2017, Klimchuk:2019a} and prominence formation \citep[\textit{e.g.},][]{Xia:2014, Keppens:2014, Xia:2016, Fan:2017, Fan:2018, Fan:2020}. In contrast to these works, we will here focus on a `levitation-condensation' configuration, where \ac{TNE} cycles are completely absent, since there will be no connectivity to lower layers, and everything is happening due to \ac{TI} and other \ac{MHD} processes. It should be noted that the pure \ac{TI} mechanism has also been studied in more idealized settings by \citet{Claes:2019,Claes:2020}, pointing out that the initial stages of the \ac{TI} process do not lead to a perfect alignment between condensations and magnetic field lines, bringing further complications to interpretations of observed fine-structure.

Prominences, once formed, come in a wide variety of shapes, sizes, and with dynamics understood to be driven by their formation, internal, and global evolution \citep[][]{Labrosse:2010, Mackay:2010, Vial:2015}. Structurally, stable prominences often appear as `clouds' within the solar atmosphere with internal striations aligned approximately to the radial direction \citep[\textit{e.g.},][]{Berger:2008}. In comparison with their on-disk counterparts, or `filaments\footnote{The terms prominence and filament are often used interchangably. We opt instead to maintain the use of `prominence' alone unless referring explicitly to the filament projection.}', similar striations or `threads' appear but are instead aligned predominantly with the solar surface \citep[\textit{e.g.},][]{Lin:2005}. For both the `prominence' and `filament' projections, the internal dynamics of the associated material are oriented in line with their internal structures, which are themselves conflicting despite being identical phenomena \citep[\textit{e.g.},][]{Zirker:1998,Berger:2008}. Furthermore, the observed magnetic field within prominences favours the picture laid out by general filament observations in that their associated magnetic field, responsible for their suspension and therein their appearance, is oriented predominantly horizontal to the solar surface \citep[as predicted by \citet{Kippenhahn:1957}, see \textit{e.g.},][]{Casini:2003,Merenda:2007,Orozcosuarez:2014,Wang:2020}.

For more than a decade now, this discrepancy between the horizonal field of a prominence and their vertical internal structuring and dynamics has been interpreted as a form of a magneto-convection process \citep[][]{Berger:2008, Berger:2010, Berger:2011, Liu:2012a}. Analysis of prominence observatons provided by the \ac{SOT} instrument on board the \ac{hinode} spacecraft led \citet{Ryutova:2010} to interpret the vertical structuring as combinations of the `plumes' and `fingers' associated with the \ac{RTI}. Numerous authors have successfully explored this interpretation further with both observations, numerical models, and combinations therein \citep[\textit{e.g.},][]{Hillier:2011, Hillier:2012a,Hillier:2012b,Keppens:2015,Kaneko:2018}. However, \citet{Low:2012a} presented an alternative theory, where this observed behaviour is instead a consequence of the quasi-periodic breakdown in the frozen-in condition, allowing prominence plasma to slip across field lines towards the surface \citep[see also][]{Low:2012b,Low:2014}. In this paper, we present the first numerical demonstration where this slippage is indeed occuring, and where the vertical motions found are not directly related to the RTI. By restricting our setup to 2.5D, true interchange or RTI activity is strongly inhibited by means of the in-plane magnetic field, while we can make direct contact with advanced linear MHD stability criteria~\citep{Goedbloed:2019}.

Finally, pioneering 3D numerical simulations of \citet{Xia:2016} effectively demonstrated the ability of condensations to spontaneously form throughout a prominence-hosting flux rope. The subsequent and complex evolution of these plasma packets to lower heights both within and out-of the global flux rope, reproduced the complex absorption and emission evolution observed within prominences as seen by the \ac{AIA} instrument on board the \ac{SDO} spacecraft \citep[cf.][]{Regnier:2011}. Here, the observed dynamics are a consequence of the condensation formation process. In this paper, we perform extreme resolution 2.5D simulations, where we are able to identify a variety of both ideal and non-ideal processes to explain similarly complex prominence-internal structure and dynamics.

\section{Numerical setup}\label{s:num_set}
To explore the small-scale structure and dynamics involved in the formation of solar prominences, we perform a series of 2.5 dimensional (2D + invariance in the 3rd dimension) MHD simulations using the parallelised \ac{AMRVAC} toolkit. The domain adopts a cartesian $(x,y)$ grid, ignoring the curvature of the Sun, oriented such that the solar surface is positioned along the $x$ axis at $y=0$. The flux rope hosting the prominence is dynamically formed using the common technique of converging motions imposed within the bottom boundary so as to approximate the cumulative effect of magnetic diffusion driven by photospheric granulation \citep[][]{vanDriel:2003}. In order to achieve the formation of cool condensations \textit{ab initio}, we include the effects of optically-thin radiative cooling, field-aligned thermal conduction, and gravity \citep[see][for the full description of how \ac{AMRVAC} handles parallel and perpendicular thermal conduction]{Xia:2018}. 

This manuscript is split largely into two separate studies. We begin in Section~\ref{ss:lowres} by expanding on the ab-initio prominence formation study of \citet{Kaneko:2015}. In Section~\ref{ss:highres}, we make full use of the \ac{AMR} capabilities of \ac{AMRVAC} to explore the dynamics of prominence condensations at scales an order of magnitude smaller than ever before. Within this study, we also vary the strength of the background magnetic field so as to explore the resulting effect on prominence formation and evolution.

\subsection{Governing equations} \label{ss:gov_eq}
The \ac{AMRVAC} toolkit is used to solve the conservative form (plasma density, momentum density, energy, and magnetic field) of the following `primitive'\footnote{MPI-AMRVAC accepts inputs in either `primitive' form \textit{i.e.}, velocity and pressure, or `conservative' form \textit{i.e.}, momentum and energy.} system of non-linear, non-adiabatic \ac{MHD} equations:
\begin{align}
	&\frac{\partial \rho}{\partial t}+\vec{\nabla} \cdot(\rho \textbf{v})=0,\label{eq:masscont}\\ 
	&\rho \frac{\partial \textbf{v}}{\partial t}+\rho \textbf{v} \cdot \vec{\nabla} \textbf{v}+\vec{\nabla} p-(\vec{\nabla} \times \textbf{B}) \times \textbf{B}-\rho \textbf{g}= 0,\label{momemtumcont}\\ 
	&\rho \frac{\partial T}{\partial t}+\rho \textbf{v} \cdot \vec{\nabla} T+(\gamma-1) [ p \vec{\nabla} \cdot \textbf{v}+\rho \textbf{g}\cdot \textbf{v}+\rho \mathcal{Q} \nonumber\\ 
	&\quad-\eta \vec{J}^2- \vec{\nabla} \cdot(\vec{\kappa} \cdot \vec{\nabla} T)]=0, \label{eq:econt}\\ 
	&\frac{\partial \textbf{B}}{\partial t}-\vec{\nabla} \times(\textbf{v} \times \textbf{B})+ \vec{\nabla} \times\eta \textbf{J}=0,\,\,\,\,\,\, \vec{\nabla} \cdot \textbf{B} =0,\label{eq:induction}
\end{align}
where $\rho, \textbf{v}, p$, and $\textbf{B}$ are the `primitive' variables for plasma density, velocity, pressure, and  magnetic field, respectively, and bold font indicates a vector quantity of three components. The final primitive variable, $T$, is the plasma temperature defined with,
\begin{equation}
	p \mu = \mathcal{R} \rho T,\label{eq:state}
\end{equation}
where $\mathcal{R}$ and $\mu$ are the gas constant and mean molecular mass, respectively, thus closing equations \ref{eq:masscont}\,--\,\ref{eq:induction}. $\gamma$ is the ratio of specific heats, under the assumption of a monoatomic ideal gas this is taken as 5/3. To account for both hydrogen and helium populations, the simulated plasma is considered fully ionised with an assumed abundance ratio of 10:1 (H:He) leading to the relation $\rho=1.4m_\mathrm{p}n_\mathrm{H}$, where $m_\mathrm{p}$ is the proton mass and $n_\mathrm{H}$ the number density of hydrogen. $\rho \mathcal{Q}$ then represents the combined effect of optically-thin radiative losses $\mathcal{L}$ and background heating $\mathcal{H}$ as,
\begin{equation}
	\rho \mathcal{Q}=\underbrace{\rho^2\Lambda(T)}_\mathcal{\rho L}-\underbrace{\rho_0^2\Lambda(T_0)e^{-2y/a}}_\mathcal{H},\label{eq:Q}
\end{equation}
where $\Lambda(T)$ is the radiative loss per unit mass as a function of temperature, $a$ the vertical magnetic scale height of the simulation, and $\rho_0$,$T_0$ are the initial density and temperature set at the bottom of the simulation, respectively, to ensure the system is initiated in thermodynamic equilibrium. This formulation implicitly assumes that the contribution of coronal heating to the solar atmosphere is an exponential with the maximum taken at photospheric heights \citep[\textit{e.g.},][]{Fang:2013,Xia:2016}. $\vec{\kappa}$ is the thermal conductivity, typically taken as the Spitzer conductivity \citep[][]{Spitzer:2006} equal to $\kappa_\parallel$=8~$\times$~10$^{-7}T^{5/2}$ erg cm$^{-1}$ s$^{-1}$ K$^{-1}$ (note that for solar coronal conditions, $\kappa_\perp$ may be considered comparatively negligible and therefore neglected from the computation). Finally, $\eta$ represents the resistivity of the system and is assumed constant across the simulation domain.

\subsection{Initial conditions}\label{ss:init_con}
The 2.5D simulation domain is initialised at $t=0$ with dimensions $-12\le x\le+12$~Mm and $0\le y\le 25$~Mm. The density distribution across the domain is stratified according to gravity i.e., hydrostatic equilibrium, with,
\begin{equation}
g(y)=g_0(r_\odot/(r_\odot+y))^2,
\end{equation}
 and the magnitude of gravitational acceleration at the solar surface $g_0$ set to 27400 cm s$^{-2}$. The density therefore decreases exponentially with height, from a maximum value of 3.2~$\times$~10$^{-15}$~g~cm$^{-3}$ at $-12\le x\le+12$, $y=0$. The background temperature of the stratified atmosphere is set to a constant value of 1~MK. As such, the pressure stratification is fixed according to the equation of state (\ref{eq:state}). Initially, the $(v_\mathrm{x},v_\mathrm{y},v_\mathrm{z})$ components of the velocity are set to 0.

Following \citet{Kaneko:2015}, we adopt an initial linear force-free background magnetic field distribution of,
\begin{align}
B_{x}&=-\left(\frac{2 L_{a}}{\pi a}\right) B_{a} \cos \left(\frac{\pi}{2 L_{a}} x\right) \exp \left[-\frac{y}{a}\right],\label{eq:bx}\\
B_{y}&=B_{a} \sin \left(\frac{\pi}{2 L_{a}} x\right) \exp \left[-\frac{y}{a}\right],\label{eq:by}\\
B_{z}&=-\sqrt{1-\left(\frac{2 L_{a}}{\pi a}\right)^{2}} B_{a} \cos \left(\frac{\pi}{2 L_{a}} x\right) \exp \left[-\frac{y}{a}\right],\label{eq:bz}
\end{align}
where $L_a$, $a$, and $B_\mathrm{a}$ are the lateral and vertical scale height, and photospheric magnitude of the linear force free field, respectively. In all of the cases that follow, $L_a=12$~Mm and $a\approx50$~Mm. Finally, in line with the simulations of \citet{Zhao:2017}, $\eta$ is set in dimensionless units at a constant value of 0.002 across the whole domain. This choice of initial conditions ensures the simulation begins in a state of near-equilibrium.

\begin{figure*}
	\centerline{\includegraphics[width=1\textwidth,clip=, trim=80 228 50 220]{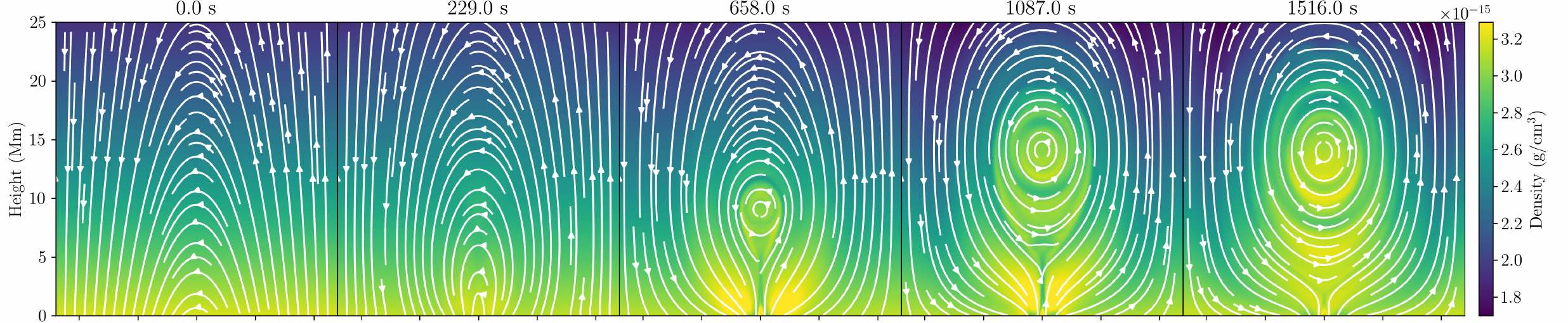}}
	\centerline{\includegraphics[width=1\textwidth,clip=, trim=80 210 50 228]{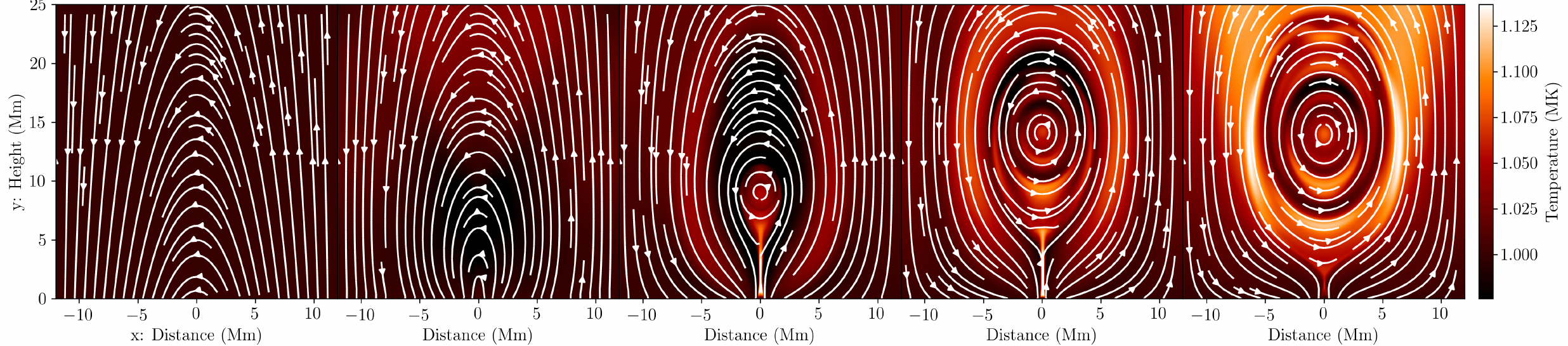}}
	\caption{The distribution of plasma density (top row) and temperature (bottom row) within the forming flux rope. Magnetic field lines have been overplotted in white to outline the isolated flux region commonly referred to as a magnetic flux rope.}
	\label{fig:den_evol_low_res}
\end{figure*}

\subsection{Boundary conditions}\label{ss:bound_con}
The boundary behaviour is set for the primitive (and converted to conservative) variables as described in table~\ref{table:bound_con}.

\begin{table}[h]
	\centering
	\begin{tabular}{l|c|c|c|c}
		\hline
		 & Left & Right & Bottom & Top \\
		\hline
		\hline
		$\rho$& 'symm' & 'symm' & $\rho_\mathrm{0}(y=0)$ & $\rho_\mathrm{0}(y=y_\mathrm{max})$\\
		$v_\mathrm{x}$& 'asymm' & 'asymm' & time dependent & 'asymm'\\
		$v_\mathrm{y}$& 'symm' & 'symm' & time dependent & 'asymm'\\
		$v_\mathrm{z}$& 'asymm' & 'asymm' & time dependent & 'asymm'\\
		$p$& 'symm' & 'symm' & $p_\mathrm{0}(y=0)$ & $p_\mathrm{0}(y=y_\mathrm{max})$\\
		$B_\mathrm{x}$& 'asymm' & 'asymm' & extrapolation & extrapolation\\
		$B_\mathrm{y}$& 'symm' & 'symm' & extrapolation & extrapolation\\
		$B_\mathrm{z}$& 'asymm' & 'asymm' & extrapolation & extrapolation\\
		\hline
	\end{tabular}
	\caption{Boundary conditions}
	\label{table:bound_con}
\end{table}
Cases 'symm' and 'asymm' refer to symmetric and asymmetric boundary conditions, respectively, which use mirror reflections about the boundary to fill the ghost cells. The $\mathrm{<}\rho/p\mathrm{>}_\mathrm{0}(y=0)$ and $(y=y_\mathrm{max})$ cases describe the enforcement of the initial values. The extrapolation indicates the use of a linear, second-order, zero-gradient extrapolation method,
\begin{align}
\mathrm{Bottom:}&~u\left( {{x_i}} \right) = \left( { - u\left( {{x_{i + 2}}} \right) + 4u\left( {{x_{i + 1}}} \right)} \right)/3,\\
\mathrm{Top:}&~u\left( {{x_i}} \right) = \left( { - u\left( {{x_{i - 2}}} \right) + 4u\left( {{x_{i - 1}}} \right)} \right)/3.
\end{align}

Following \citet{Kaneko:2015} and \citet{Zhao:2017}, the construction of the flux rope topology from the initial linear force-free field is achieved through a combination of converging and shearing footpoint motions imposed on the bottom boundary, and magnetic reconnection controlled using the numerical diffusion $\eta$ parameter of eq.~(\ref{eq:induction}). The former is prescribed according to observational theory \citep[\textit{e.g.},][]{vanBallegooijen:1989} and practically, here, in a similar way as \citet{Kaneko:2015},
\vspace{-0.1cm}
\begin{equation}
	\begin{tabularx}{0.5\textwidth}{c|c|c}
		& $~~\left(0 < x<L_{a} / 4, y<0\right)~~$&$~~\left(L_{a} / 4 \leqslant x \leqslant L_{a}, y<0\right)~~$\\
		\hline
		$\left\{\begin{matrix}
				v_\mathrm{x}\\
				v_\mathrm{y}\\
				v_\mathrm{z}
			\end{matrix}\right\}=$&
			$\left\{\begin{matrix}
				-v_{0}(t) \frac{x}{L_{a} / 4}\\
				0\\
				v_{0}(t) \frac{x}{L_{a} / 4}
			\end{matrix}\right\}$&
			$\left\{\begin{matrix}
					-v_{0}(t) \frac{L_{a}-x}{3L_{a} / 4}\\
					0\\
					v_{0}(t) \frac{L_{a}-x}{3L_{a} / 4}
				\end{matrix}\right\}$
	\end{tabularx}
\end{equation}
\vspace{-0.7cm}
\begin{equation}
\begin{tabularx}{0.5\textwidth}{c|c|c}
& $\left(-L_{a} / 4 < x<0, y<0\right))$&$\left(-L_{a}\leqslant x \leqslant -L_{a} / 4, y<0\right)$\\
\hline
$\left\{\begin{matrix}
v_\mathrm{x}\\
v_\mathrm{y}\\
v_\mathrm{z}
\end{matrix}\right\}=$&
$\left\{\begin{matrix}
v_{0}(t) \frac{|x|}{L_{a} / 4}\\
0\\
-v_{0}(t) \frac{|x|}{L_{a} / 4}
\end{matrix}\right\}$&
$\left\{\begin{matrix}
v_{0}(t) \frac{L_{a}-|x|}{3L_{a} / 4}\\
0\\
-v_{0}(t) \frac{L_{a}-|x|}{3L_{a} / 4}
\end{matrix}\right\}$
\end{tabularx}
\end{equation}
with the velocity $v_{0}(t)$ a function of time,
\begin{alignat}{3}
v_{0}(t)&=v_{00}, &&\quad\left(0<t<t_{1}\right),\\
v_{0}(t)&=v_{00}\left(t_{2}-t\right) /\left(t_{2}-t_{1}\right), &&\quad\left(t_{1} \leqslant t<t_{2}\right),\\
v_{0}(t)&=0, &&\quad\left(t \geqslant t_{2}\right),
\end{alignat}
where $t$, $t_{1}$, and $t_{2}$ are the current time, and the time the converging and shearing velocity begins to slow ($t_1=1000$~s) and stop ($t_2=1500$~s), respectively, with $v_{00}=12$~km~s$^{-1}$. The near-equilibrium state achieved in the initial conditions is permitted a short relaxation period of $\approx$~200 s to ensure that any initial relaxation effects are smoothed out before the driving motions at the boundary are imposed. Time $t=0$ is defined at the moment these driving motions are initiated.

\subsection{Numerical methods}\label{ss:num_met}

The equations (\ref{eq:masscont})\,--\,(\ref{eq:induction}) are solved using a three-step, third-order Runga-Kutta method, with a HLL approximate Riemann solver making use of the third-order slope limiter presented by \citet{Cada:2009}. In our explicit time-stepping procedure, we adopt a courant value of 0.8. To negate the difficulties in handling large magnetic field strengths, \ac{AMRVAC} permits the treating of the magnetic field in a split way following \citet{Tanaka:1994}, but extended to the linear force free case as explained in \citet{Xia:2018}, where the analytic equilibrium current also enters. As such, the magnetic field of equations (\ref{eq:bx})\,--\,(\ref{eq:bz}) is specified as a time-independent background field, and we evolve the deviation from this field in a fully nonlinear MHD evolution. $\vec{\nabla} \cdot \textbf{B}=0$ is ensured on an AMR grid by using the diffusive approach as explained in \citet{Keppens:2003}, wherein a source term is added in a split manner to the energy equation (\ref{eq:econt}) to smoothly diffuse/advect any present monopole errors and reduce the signature of $0\lessapprox \vec{\nabla} \cdot \textbf{B} \ll 1$. Note that \ac{AMRVAC} has many different means to control monopole errors discretely, such as by using source terms, or by solving an elliptic projection equation \citep[see][]{Teunissen:2019}, or by exploiting a constrained transport approach.

\begin{figure}
	\centerline{\includegraphics[width=0.5\textwidth,clip=, trim=110 130 85 140]{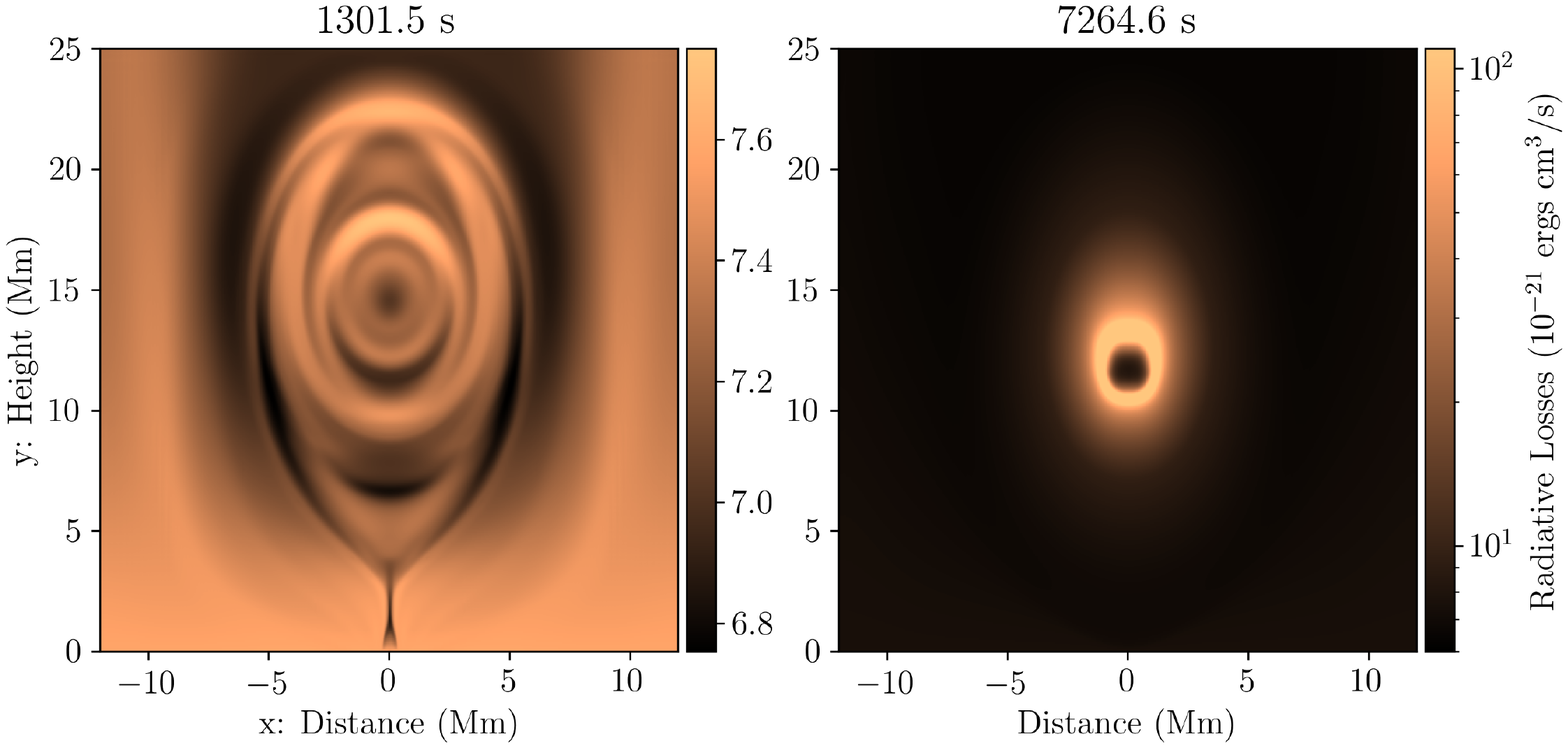}}
	\caption{A comparison between both the distribution and the magnitude of radiative losses within the flux rope \textit{left}; immediately after the formation of the flux rope and \textit{right}; just prior to the appearance of condensing material.}
	\label{fig:rad_evol}
\end{figure}

%
%

\begin{figure*}
	\centerline{\includegraphics[width=1\textwidth,clip=, trim=80 244 30 220]{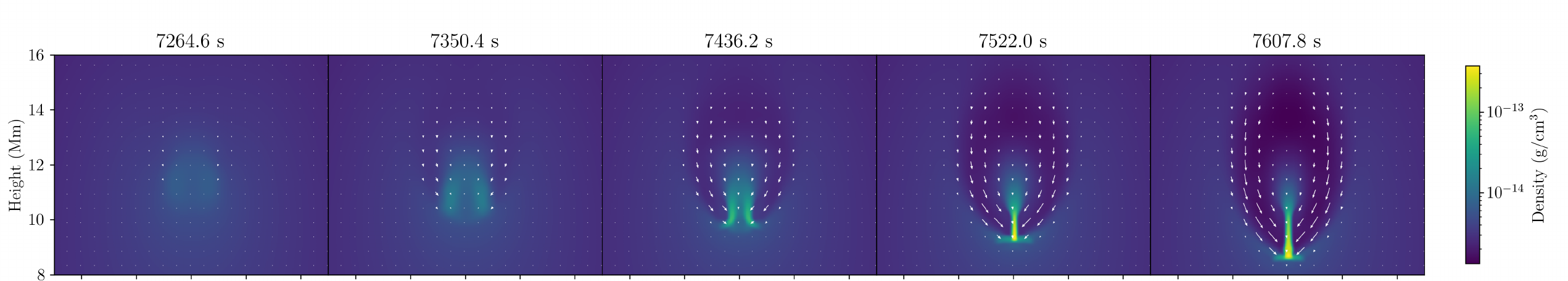}}
	\centerline{\includegraphics[width=1\textwidth,clip=, trim=80 244 30 243]{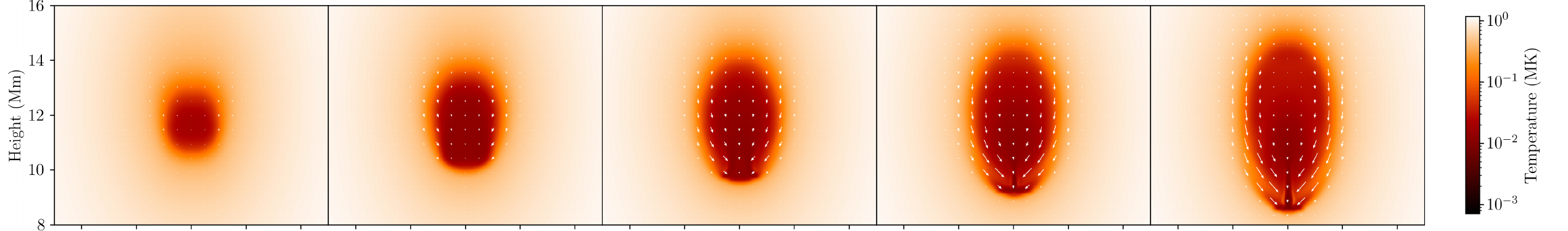}}
	\centerline{\includegraphics[width=1\textwidth,clip=, trim=80 244 30 243]{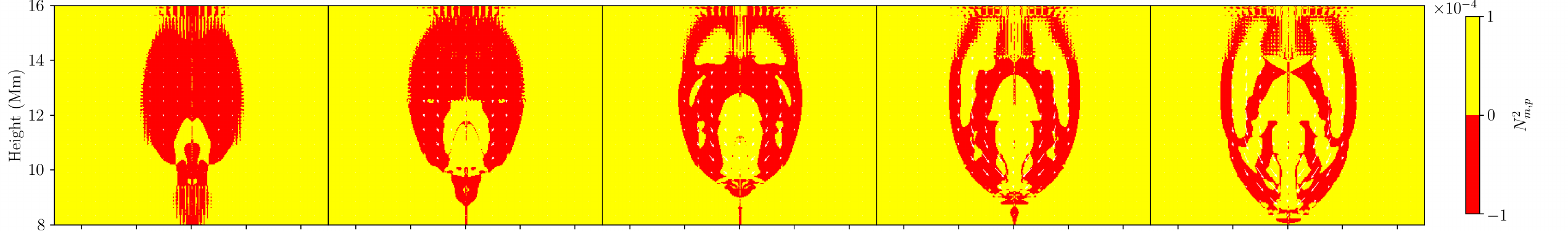}}
	\centerline{\includegraphics[width=1\textwidth,clip=, trim=80 220 30 243]{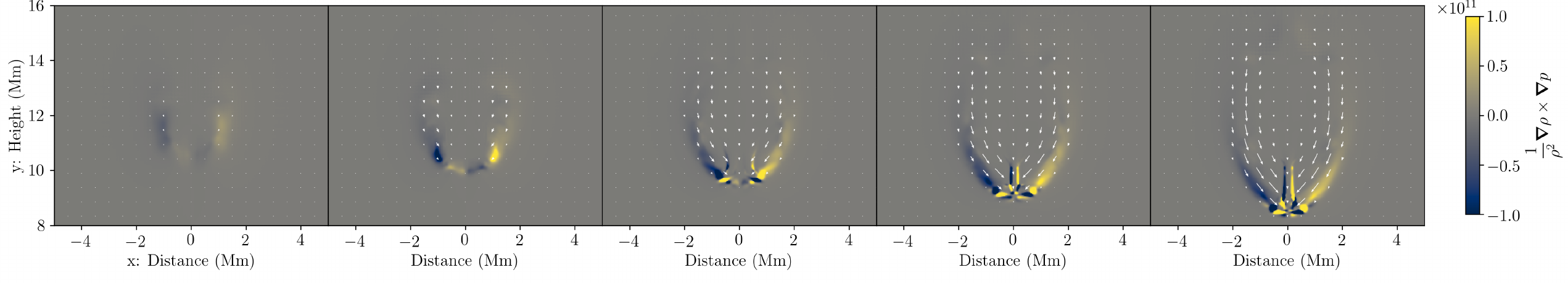}}
	\caption{Evolution of the condensing material within the formed flux rope of the 31~km resolution case. The \textit{first} row details the evolution of the density, \textit{second}; temperature, \textit{third}: Brunt-V\"ais\"al\"a instability ($N_{m, p}^{2}$), and \textit{fourth}; baroclinicity (positive $\rightarrow$ clockwise rotation). Velocity quivers of relative magnitude are overplotted as white arrows to indicate the direction of, and increase in, condensation rate with time. A movie of this figure is available with the online version of this manuscript.}
	\label{fig:condensation_evol_low_res}
\end{figure*}

\section{Analysis  methods and results}\label{s:results}
\subsection{Baseline  31~km resolution simulation}\label{ss:lowres}

To establish a baseline for comparison between this work and that of previous authors, the first aim of this study is to compare and contrast the results of \citet{Kaneko:2015} with the one obtained using \ac{AMRVAC} under the numerical schemes described in the preceeding sections. To that end, we adopt a grid of 96~$\times$~96 with 3 additional levels of refinement permitting a maximum effective grid size of $\approx$~31~km \citep[approximately equal to the highest resolution case explored by][]{Kaneko:2015}. We also adopt the same piecewise radiative cooling curve of \citet{Hildner:1974}. We deviate from the \citet{Kaneko:2015} work in two main aspects (1) the exponential heating function of eq.~(\ref{eq:Q}) is chosen in line with our recent works for consistency \citep[\textit{e.g.},][]{Fang:2013,Xia:2016,Zhao:2017}, (2) and we remove the minimum temperature approximation that was used in their study (likely for numerical stability reasons). The strength of the initial surface magnetic field is set to 3~G, as in \citet{Kaneko:2015}.

\subsubsection{Global evolution}\label{sss:global_ev_lowres}

Figure~\ref{fig:den_evol_low_res} presents the evolution of the density, temperature, and magnetic field during the `build phase' of the magnetic flux rope. Material is elevated from the bottom of the `atmosphere' and collects within the flux rope alongside the reconnection process. Here, the current sheet coincides with a decrease in density and an increase in temperature at the neutral line ($x=0$). The persistence of an inverted `Y' configuration below the forming flux rope enables reconnection to continue to supply both plasma and flux. By the time the footpoint motions have stopped and the flux rope has formed, we see a clear difference in the distribution of temperature and density at $t\approx~1500$~s in comparison with $t=0$.

After the cessation of the converging boundary motions at $t=1300$~s, the system enters a relaxation phase where the flux rope that was accelerated upwards through the atmosphere undergoes damped periodic oscillations. The system remains in this phase for approximately 6200~s. Figure~\ref{fig:rad_evol} details how the optically-thin radiative losses $\Lambda(T)$ evolve during this time, indicating a strong increase in emission around 7000~s as expected when the material within the flux rope begins to cool. It should be noted that the cooling process is linked directly to the linear thermal instability \citep[cf.][]{Field:1965, Xia:2014, Xia:2016, Claes:2019, Claes:2020}, as in this scenario there is no TNE cycle in our 2.5D simulation.

Shortly thereafter, and as shown in the top row of Figure~\ref{fig:condensation_evol_low_res}, a large condensation begins to form in the wake of the increased radiative losses, the associated material has an initial velocity of $\approx$~30~km~s$^{-1}$ at $t=7264.6$~s. Around $7264.6<t<7350.4$~s, the shape and density of the condensation is similar to that described by \citet{Kaneko:2015}. In contrast to this work, we find a subsequent collapse of this dense cloud leading to smaller-scale structure. This difference can be ascribed to the artificial and fairly high temperature cut-off adopted by these authors. Indeed, in Figure~\ref{fig:condensation_evol_low_res} much lower temperatures of order $<10^4$~K are recorded within the forming condensations. As time progresses, we see the condensation narrowing at a velocity of $\approx$~80~km~s$^{-1}$ to form a single, vertically-oriented monolithic structure. This is reminiscent of the assumptions adopted in many \ac{RTE} inversion methods \citep[\textit{e.g.},][]{Beckers:1964,Heinzel:2001,Asensioramos:2008, Asensioramos:2011,Schwartz:2019}. The velocity of the material involved in the condensation process is up to an order of magnitude greater than that observed within post-formation prominence condensations \citep[cf.][]{Labrosse:2010}, but of the same order of magnitude as previous, fully 3D simulations of this process \citep[\textit{e.g.},][]{Xia:2016}.

\subsubsection{The formation of a `monolithic' prominence}\label{sss:form_mon_lowres}


In the presence of a gravitational field, plasma can fall in the direction of decreasing gravitational potential. A gravitationally stratified atmosphere in hydrostatic equilibrium will have a density stratification that decreases exponentially with a scale height equal to the pressure scale height $H$,
\begin{equation}
	H= \frac{k_\mathrm{B}T}{\mu g},\label{eq:hdequilibrium}
\end{equation}
where $k_\mathrm{B}$ is the Boltzmann constant, $T$ the gas temperature, $\mu$ the mean molecular mass, and $g$ the acceleration due to gravity. A small packet of this gas displaced out of hydrostatic equilibrium is then said to oscillate at the \ac{BV} frequency. The presence of damping  will then return this displaced packet of gas to equilibrium.

In the presence of a magnetic field, there is the additional magnetic pressure (+ tension) that, depending on the magnetic topology, may levitate charged gas to altitudes far in excess of the gas pressure scale height \citep[cf.][]{Hillier:2013}. To be explicit, the magnetic force $\vec{F_\mathrm{m}}$ supplied to ionised gas (plasma) may be decomposed as follows,
\begin{equation}
\vec{F_\mathrm{m}}=\textbf{J}\times \textbf{B}=(\vec{\nabla} \times \textbf{B}) \times \textbf{B}=\underbrace{-\nabla\left(\frac{B^{2}}{2}\right)}_{B_\mathrm{pressure}}+\underbrace{(\vec{B} \cdot \vec{\nabla}) \mathbf{B}}_{B_\mathrm{tension}},
\end{equation}
wherein,
\begin{equation}
B_\mathrm{tension}=B(\hat{\vec{b}} \cdot \vec{\nabla})(B \hat{\mathbf{b}})=\hat{\mathbf{b}} \hat{\mathbf{b}} \cdot \nabla\left(\frac{B^{2}}{2}\right)+B^{2} \frac{\hat{\mathbf{n}}}{R_\mathrm{c}},
\end{equation}
and hence magnetic pressure acts in \textit{all} directions whereas magnetic tension acts to nullify the parallel force, and has a component only in the direction $\hat{\textbf{n}}$ of the radius of curvature $R_\mathrm{c}$.  Note we have chosen units such that the value of the permeability of free space $\mu_{0}=1$.

\citet{Blokland:2011a} studied the application of this to prominences in hydrostatic equilibrium that represent the flux-rope embedded prominence stage that we reached in this simulation. Indeed, the 2.5D nature of the problem has nicely nested poloidal flux surfaces that can levitate plasma above the solar surface by magnetic pressure and tension. After the prominence forms, we reach a phase of much slower evolution, and this can be seen as a succession of 2.5D magnetostatic equilibria with prominence plasma nested along flux surfaces.


\citet{Blokland:2011b} went on to analyse these perfectly force-balanced states, in terms of linear MHD stability. A crucial factor in this linear stability analysis is played by the flux-surface projected forms of the \ac{BV} frequencies, given by,
\begin{align}
N_{B, p}^{2} &=\left[\frac{\vec{B} \cdot \vec{\nabla} p}{\rho B}\right]\left[\frac{\vec{B}}{\rho B} \cdot\left(\vec{\nabla} \rho-\frac{\rho}{\gamma p} \vec{\nabla} p\right)\right], \label{eq:BVF}\\ 
N_{m, p}^{2} &=\left[\frac{\vec{B} \cdot \vec{\nabla} p}{\rho B}\right]\left[\frac{\vec{B}}{\rho B} \cdot\left(\vec{\nabla} \rho-\frac{\rho}{\gamma p+B^{2}} \vec{\nabla} p\right)\right],\label{eq:BVF+M}
\end{align}
where $N_{B, p}$ and $N_{m, p}$ are the traditional and magnetically-modified (\textit{i.e.}, including magnetic pressure) \ac{BV} frequencies, respectively. When (parts of) flux surfaces exist where this $N^2<0$, the force-balanced equilibrium state is found to be unstable, since the organised continuous parts of the MHD spectrum become overstable. The authors refer to this as driving a \acf{CCI}, which is possible for those magnetostatic equilibria where the density variation in the flux rope behaves as a flux function. For our simulations here, the density is indeed virtually constant on every nested flux surface for a long time during the simulation (prior to about 7000 s), as seen in Figs.~\ref{fig:den_evol_low_res}-\ref{fig:condensation_evol_low_res} and the online animations. 

\citet{Moschou:2015} studied the application of the \ac{CCI} theory to the coronal rain phenomenon within a 3D numerical simulation. The authors note that the evolution of their condensations could not be directly related to the \ac{CCI} as that setup was done in an arcade and hence had no nested flux surfaces, also due to lacking the required translational symmetry. Nevertheless, the spatial coincidence of $N^2<0$ regions with actual condensations remained consistent, concluding that the motions of the plasma were driven at all times by some variant of the \ac{CCI}. Following these authors, the third row of Figure~\ref{fig:condensation_evol_low_res} details the evolution of the magnetically-modified \ac{BV} frequency within a cutout of the simulation domain. At $t=7264.6$~s we find that the $N_{m, p}^{2}<0$ regions trace the majority of those flux surfaces on which material is condensing. The material present along these specific flux surfaces may therefore be considered \ac{CCI}-unstable. After this time we find that the $N_{m, p}^{2}<0$ signatures are highly structured and the plasma located in these $N_{m, p}^{2}<0$ regions will slide along their own magnetic surface under the influence of gravity, until the pressure balance is restored. 

\begin{figure}
	\centerline{\includegraphics[width=0.45\textwidth,clip=, trim=100 60 50 50]{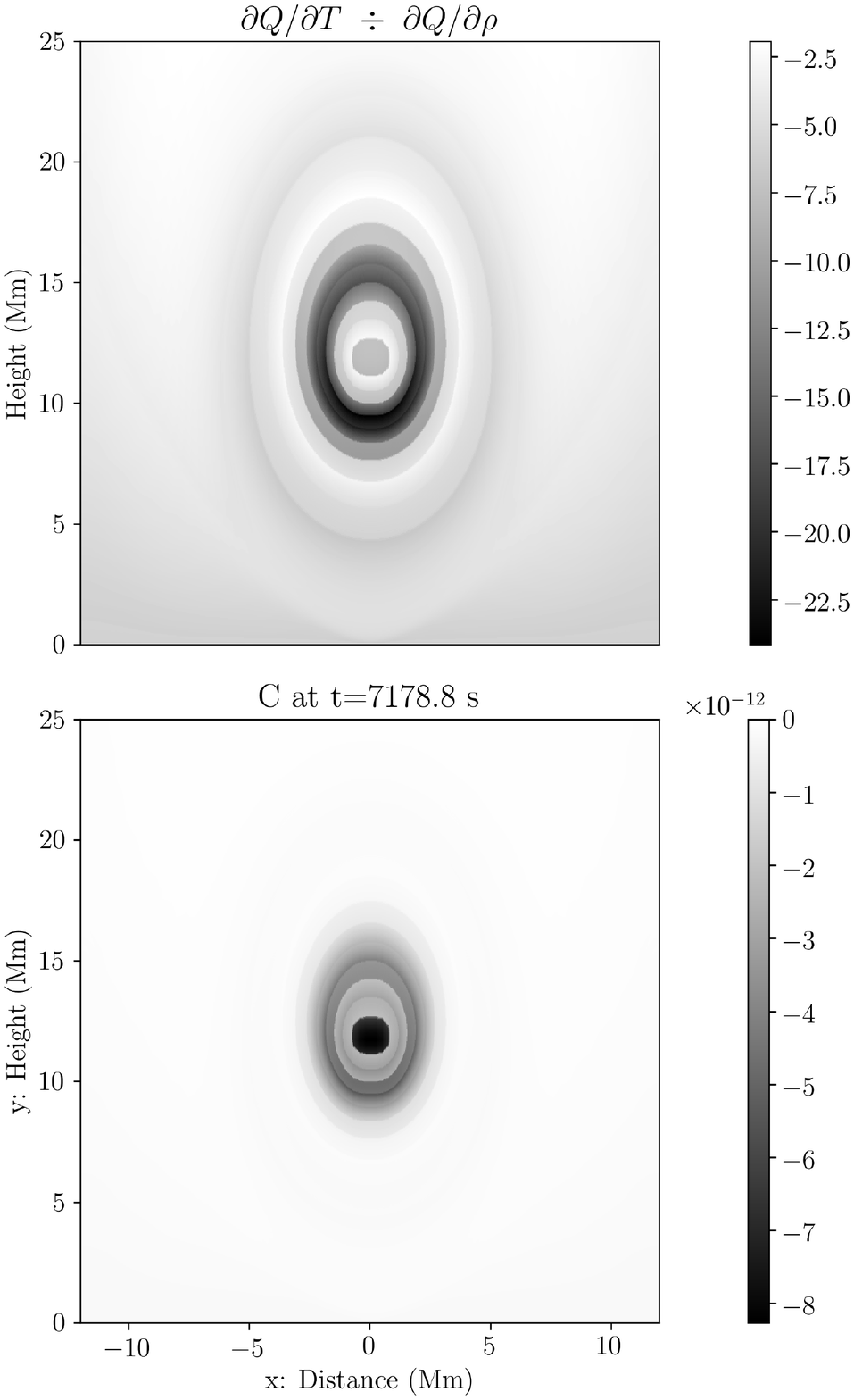}}
	\caption{Locations of thermal instability within the 31~km gridstep prominence. \textit{Top}; Ratio of thermal to density influence on the varying energy loss $\mathcal{Q}$ term in eq.~(\ref{eq:econt}). \textit{Bottom}; Spatial map of the isochoric linear stability metric $C$ (eq.~\ref{eq:C}) for the thermal mode instability.}
	\label{fig:lowres_TI}
\end{figure}

\begin{figure*}
	\centerline{\includegraphics[width=1\textwidth,clip=, trim=80 195 30 180]{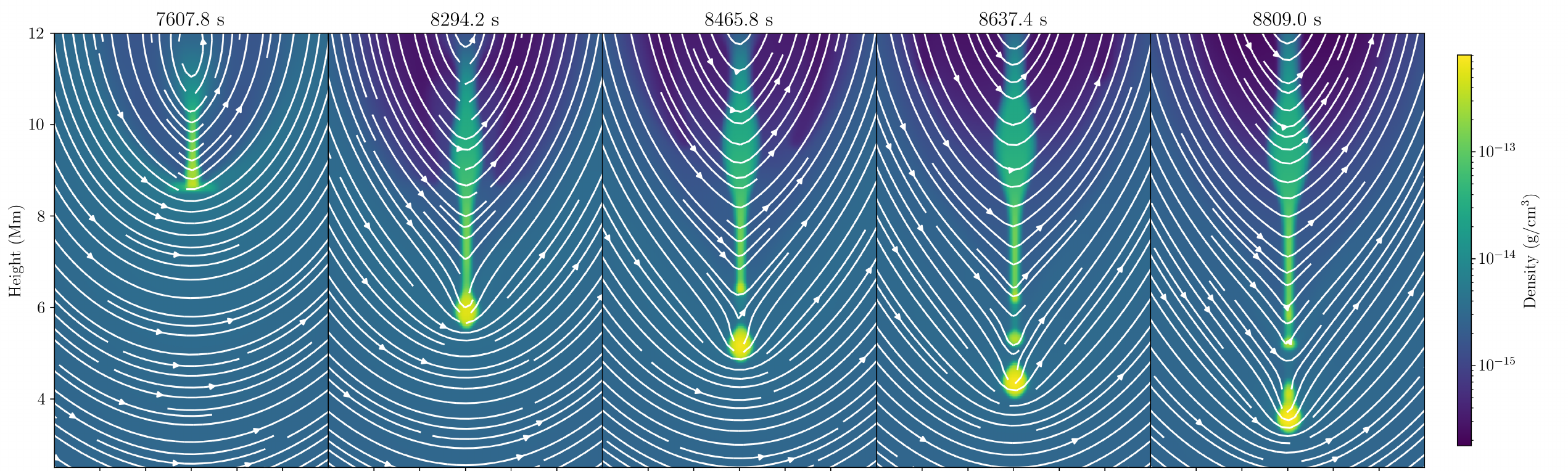}}
	\centerline{\includegraphics[width=1\textwidth,clip=, trim=80 195 30 192.5]{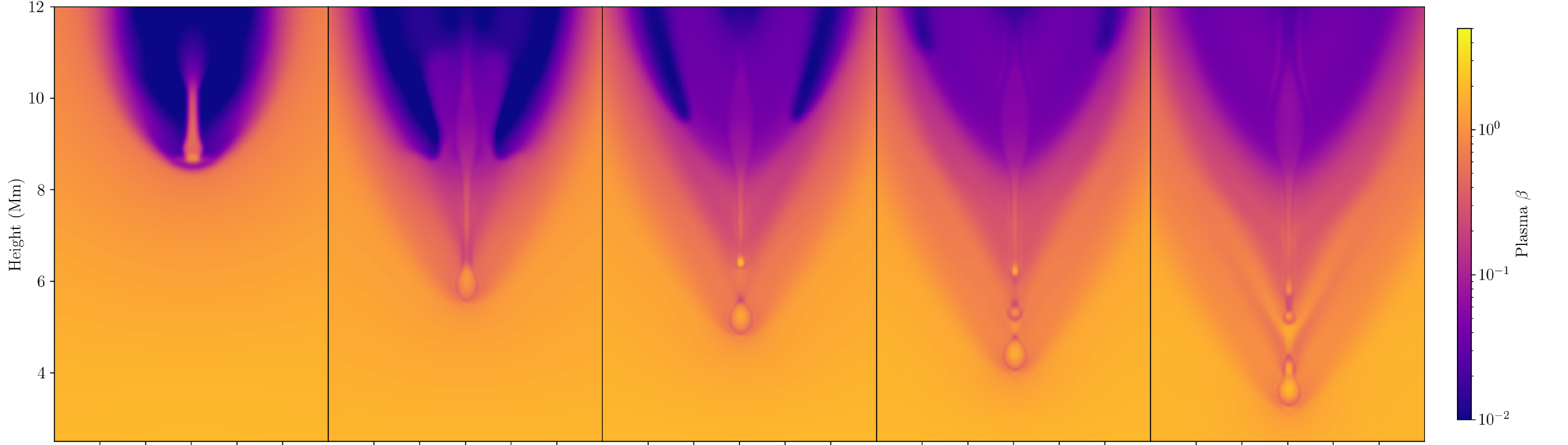}}
	\centerline{\includegraphics[width=1\textwidth,clip=, trim=80 175 30 192.5]{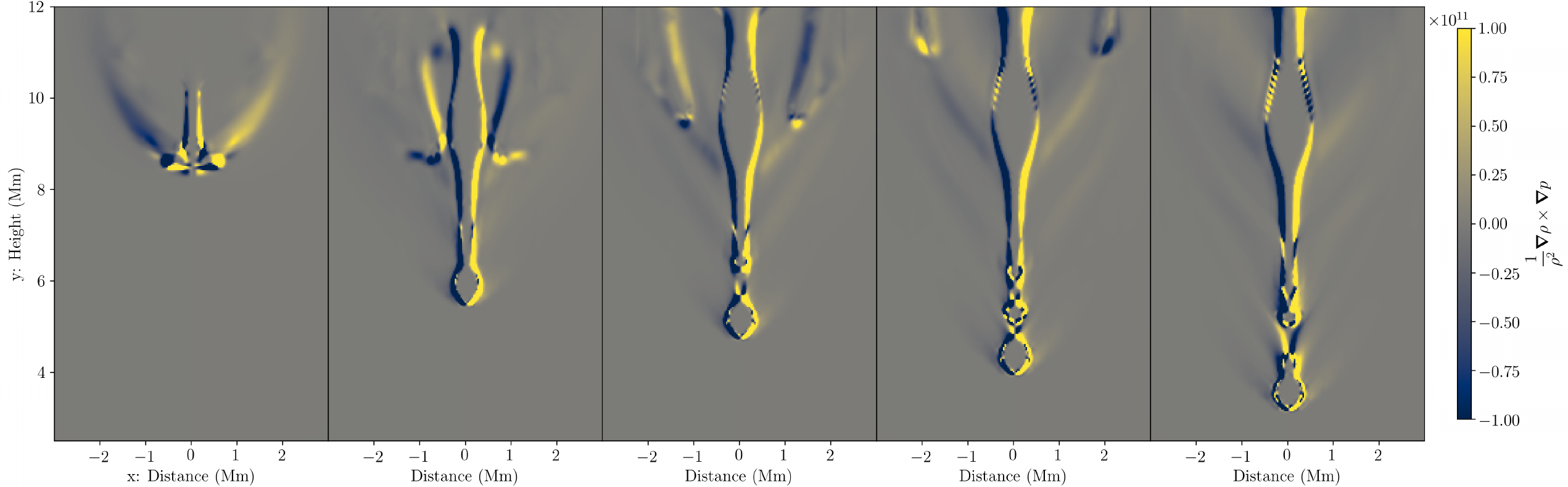}}
	\caption{The evolution of the monolithic prominence once formed by catastrophic, optically-thin radiative cooling. The \textit{top} row presents the evolution of the density with magnetic field lines indicated using white streamlines,. The \textit{middle} row details the value of plasma~$\beta$ throughout the flux rope and prominence. The \textit{bottom} row shows the evolution of the baroclinicity.}
	\label{fig:rti_evol_low_res}
\end{figure*}

During the very early stages of condensation, the plasma within a flux rope is being redistributed as a consequence of the unstable flux projected \ac{BV} distribution. At any given moment, a short-lived localisation of density may occur that increases the radiative energy losses $\rho\mathcal{L}$ in this location. As such, the associated plasma will cool. For plasma at a temperature of 1~MK, the cooling curve of \citet{Hildner:1974} indicates the radiative losses would increase as a result of this cooling, leading to a feedback process. However, for sufficiently small temperature perturbations, thermal conduction smooths out these localised variations. Hence, the (mostly temperature) gradient of the cooling curve must be steep enough in combination with a sufficiently large temperature perturbation. Furthermore, as before, the localised decrease in temperature is accompanied by a localised increase in density, where the $\rho^2$ dependence on the radiative losses then ensures this cooling process cannot be easily halted or reversed. \citet{Parker:1953} and \citet{Field:1965} derived conditions for this \ac{TI} under isochoric and isobaric conditions, respectively. \citet{Xia:2012} studied the formation of prominence condensations through the \ac{TI} and noted that the temperatures and pressures of the condensations dropped both faster and earlier than their densities, as was alluded to above. Therein they find the isochoric assumption more relevant for prominence condensations than the isobaric \citep[an identical conclusion was also reached by][]{Moschou:2015}. The \ac{TI}-unstable criteria $C$ for the isochoric assumption is given by,
\begin{equation}
C=k^{2}-\frac{1}{\kappa}\left(\frac{\partial \mathcal{H}}{\partial T}-\frac{\partial (\rho \mathcal{L})}{\partial T}\right)<0, \label{eq:C}
\end{equation}
where $k$ is the wavenumber of the perturbation and $\kappa$ is the coefficient of thermal conduction. These authors also noted the relative insensitivity of $C$ to the value of $k$ in the isochoric case. In line with previous studies, $k$ takes the value of double the condensation size ($\sim$~2~Mm). The top panel of Figure~\ref{fig:lowres_TI} demonstrates that the temperature decreases are responsible for driving the \ac{TI} evolution (bottom panel) in the early stages of the prominence studied here. We find that the localisation of the largest negative values for $C$ coincide with the locations of the condensations that form at a later time as shown in Figure~\ref{fig:condensation_evol_low_res}.

\begin{figure*}
	\centerline{\includegraphics[width=1\textwidth,clip=, trim=0 125 0 125]{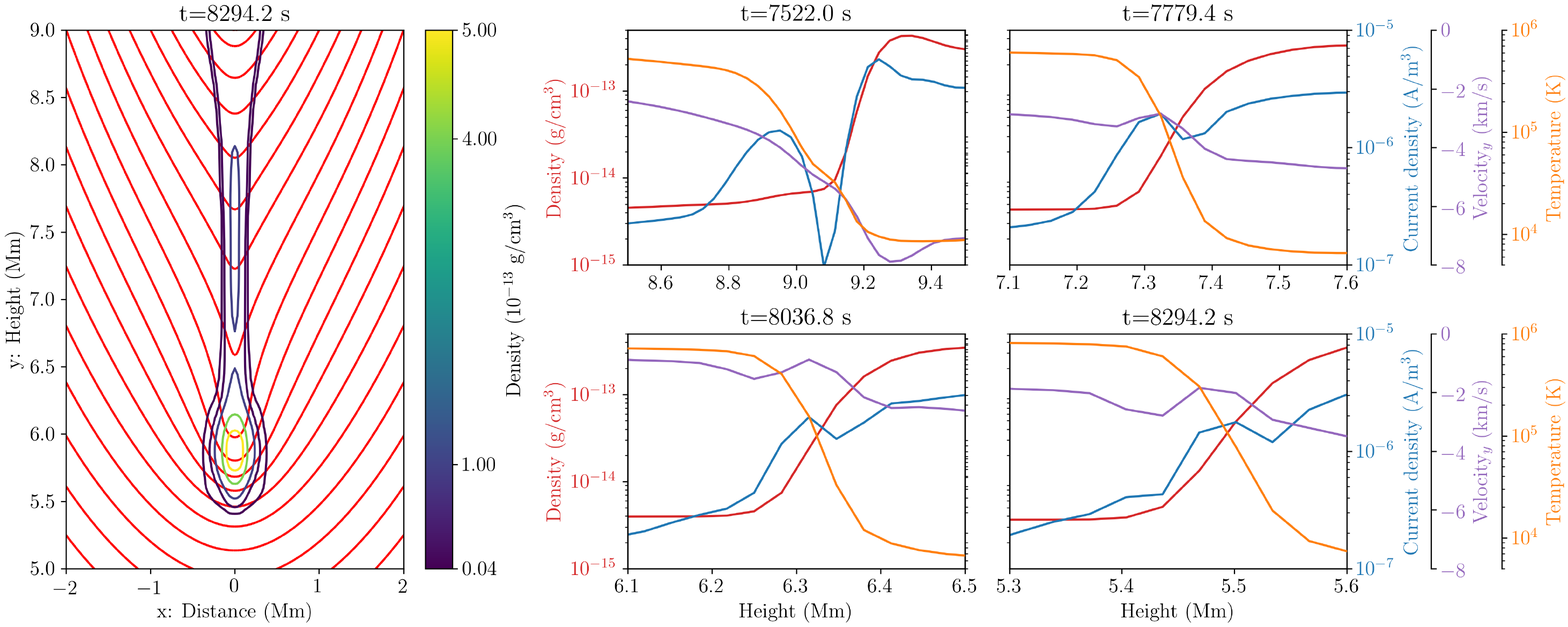}}
	\caption{Magnetic field and density map, and evolution of plasma density, current density, vertical velocity, and temperature a long a cut taken vertically through $x=0$. \textit{Left}; a zoom in of the region containing the bottom of the prominence condensation. Magnetic field lines are plotted in red with density contours overlaid according to the colorbar. \textit{Right}; four panels detailing the evolution of the aforementioned physical parameters across the leading edge of the falling condensation, times are indicated as titles to each plot.}
	\label{fig:low_2014_t}
\end{figure*}

Prior to the coalescence of the final monolithic structure, local density enhancements appear to form within the general condensation, for example see the density panel at $t=7436.2$~s. Considering the background, vertically stratified atmosphere at this stage, it is clear to see that denser material finds itself positioned \textit{above} less dense material. In hydrodynamics, such density inversions may become unstable to the \ac{RTI}. The behaviour of material related to the \ac{RTI} is then typically quantified by,
\begin{equation}
\mathcal{A}=\frac{\rho_{+}-\rho_{-}}{\rho_{+}+\rho_{-}},
\end{equation}
where $\mathcal{A}$ is the Atwood number, and $\rho_{+,-}$ is the more/less dense region, respectively. For values of $\mathcal{A}\gtrapprox0$, we find that the mixing is somewhat symmetrical and the less-dense plasma penetrates into the more-dense plasma to an equal degree as the reverse. For $\mathcal{A}\approx1$, the more-dense plasma penetrates further into the less-dense region with relatively little of the reverse \citep[][]{Chandrasekhar:1955}. \citet{Moschou:2015} explored the distribution of $\mathcal{A}$ within their 3D simulations of coronal rain, calculated as a search for a density gradient in the $y$-direction of their domain. However, the $\mathcal{A}$ parameter quantifies the density stratification in \textit{only} the $y$-direction with particular emphasis on the locations of density gradients. The motions of plasma subject to the \ac{RTI} are also dependent on the pressure gradient that the evolving plasma encounters and, in fact, the misalignment between these pressure and density gradients leads to the vortices commonly observed in the wake of the main falling `fingers' of \ac{RTI}-unstable material. However, the 2.5D configuration used here is oriented across the flux loops of the flux rope and hence these vortices are heavily stabilised by the magnetic tension within the topological dips. As such, we do not expect to observe similar large-scale structures to those recorded in previous simulations of prominence evolution involving $\vec{B}$ oriented across the 2.5D plane, or in full 3D \citep[\textit{e.g.},][]{Hillier:2012a,Hillier:2012b, Terradas:2015, Keppens:2015, Kaneko:2018,Popescubraileanu:2020}. Nevertheless, we expect the misalignments between the pressure and density gradients to be present in 2.5D and therein induce motions inline with the \ac{RTI} but on scales of the order of a flux tube. Evolution of the plasma in this way may also be characterised using baroclinicity, a component of the evolving vorticity $\vec{\omega}$ that arises in a flow with anisotropic temperature,
\begin{equation}
\frac{\partial \vec{\omega}}{\partial t}+(\textbf{v} \cdot \vec{\nabla}) \vec{\omega}=(\vec{\omega} \cdot \vec{\nabla}) \textbf{v}-\vec{\omega}(\vec{\nabla} \cdot \textbf{v})+\underbrace{\frac{1}{\rho^{2}} \vec{\nabla} \rho \times \vec{\nabla} p}_{\mathrm{baroclinicity}}.
\end{equation}
This baroclinicity considers that a misalignment between the pressure and density gradients gives rise to the rotation of plasma as it \textit{slides} down a constant-pressure gradient. In general, plasma unstable to the \ac{RTI} in a direction other than vertical may be identified using this approach. The fourth row of Figure~\ref{fig:condensation_evol_low_res} details the evolution of the baroclinicity within the condensing material. At $t=7264.6$~s, the signature of baroclinicity covers a broad region but one that is focussed at the periphery of the condensation region and forms a `U' shape. As the evolution progresses, this `U' shape is amplified in magnitude, but also becomes flat and increasingly localised to the bottom of the forming monolithic structure.

In the density panels, at $t=7436.2$~s we see that the two early condensations develop a curvature as they move towards the concave-up portions of the field topology. This is consistent with the baroclinicity-induced rotation. It's important to emphasise here that vorticity maps quantify, generally, the rotation of the velocity field (induced by some physical process), whereas the baroclinicity maps show the source of vorticity and represent a physical process in itself. Hence, maps of the vorticity would likely also show this rotation but would not offer information on the driving process.

\subsubsection{Evolution of the `monolithic' prominence}\label{sss:ev_mon_lowres}

Figure~\ref{fig:rti_evol_low_res} details the evolution of the now-formed prominence after the initial condensation period. Initially, between $7607.8<t<8294.2$~s, it can be seen that the condensation region significantly deforms the field topology of the host flux rope as the concentration of highest-density falls from an altitude of $\approx9$\,--\,$6$~Mm with an average velocity of 5~km~s$^{-1}$. The largest magnetic field strength is recorded within the condensation at this time, with a value of $\approx~4.2$~G. The falling material drags the associated field lines down and those field lines that lie directly below the condensation are compressed. Shortly after $t=8294.2$~s, reconnection is triggered just above the densest portion of the falling monolith, creating a magnetically-isolated region of high density that continues to fall, now with a velocity of $\approx$~10~km~s$^{-1}$. Behind this falling `plasmoid'\footnote{A common name attributed to a collection of plasma and magnetic field, isolated from its surrounding environment \textit{i.e.}, a magnetic island.}, another high-density region forms in its wake and at $t\approx8500$~s reconnection is again triggered behind this new region. Once the new density enhancement is isolated by the reconnection, it begins to fall much faster ($v\approx$~7~km~s$^{-1}$) and eventually approaches and coalesces with the first falling plasmoid which has slowed to $\approx$~5~km~s$^{-1}$ at this time due to increasing magnetic tension, see the $t=8809.0$~s panel of Figure~\ref{fig:condensation_evol_low_res}.

The separations of the plasmoids from the main monolithic, prominence condensation are comparable to \ac{AIA} observations reported by \citet{Liu:2012a}. The authors related the apparent evolution of dynamic prominence plasma observed in the cooler 304~\AA\ passband to the \ac{RTI}. Indeed, the simulation here and that of \citet{Kaneko:2018} contain plasma stratifications that are remarkably similar to these observations, albeit differing from the small-scale dynamics produced in dedicated simulations of the \ac{RTI} in prominences \citep[\textit{e.g.},][]{Hillier:2012b,Keppens:2015}. To explore this, the bottom row of panels in Figure~\ref{fig:rti_evol_low_res} shows the evolution ($7607.8<t<8809.0$~s) in the distribution of baroclinicity within a cut-out \ac{FOV} covering only the region of the plasmoid motions. 

As in Figure~\ref{fig:condensation_evol_low_res}, the strong signatures of baroclinicity are focussed at the edges of both the main condensation and the falling plasmoids. The predominantly negative-left and positive-right signatures on either side of the main condensation are caused by a relatively strong (weak) horizontal (vertical) pressure gradient across the condensation boundary (recognise that $\boldsymbol{\nabla}\rho \times \boldsymbol{\nabla}p=[\partial_x\rho\partial_y p-\partial_y\rho \partial_x p] \mathbf{e}_z$, so we always show the out-of-plane component in figures). The same is true for the left- and right-sides of the falling plasmoids, the relative importance of this is discussed in more detail in Section~\ref{s:discussion}.

\begin{figure*}
	\centerline{\includegraphics[width=1\textwidth,clip=, trim=0 100 0 100]{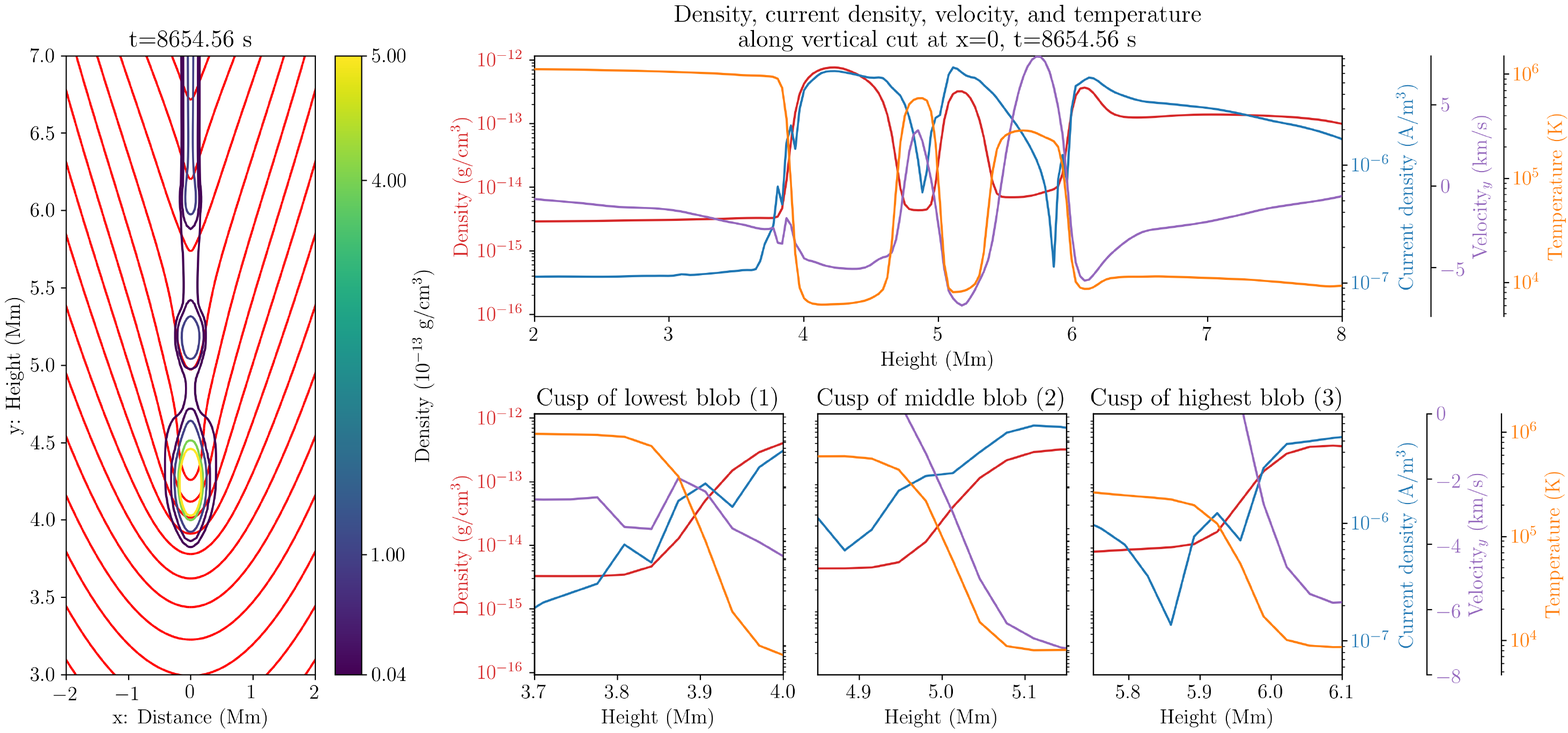}}
	\caption{Magnetic field and density map, and stratification of plasma density, current density, vertical velocity, and temperature along a cut taken vertically through $x=0$. \textit{Left}; a zoom in of the region containing the multiple prominence condensations. Magnetic field lines are plotted in red with density contours overlaid according to the colorbar. \textit{Top}; the stratification of the aforementioned plasma parameters between 2\,--\,8~Mm, isolating the three prominence consensations. \textit{Bottom}; Three panels focussing on the distribution of parameters across the cusp of each blob.}
	\label{fig:low_2014_3}
\end{figure*}

According to \citet{Low:2012a}, the initial compression of the magnetic field by the heavy prominence should create a large but finite current between the flux surfaces that, through dissipation, allows material to pass from one field line to another in the direction of gravity (cf. their Figure~8). Hence, material may slip across field lines,  a theory invoked to offer a potential explanation to why quiescent prominences projected above the limb appear to have a more-vertical structuring than their on-disk, seemingly-horizontal counterparts \citep[see also][]{Low:2012b}. \citet{Low:2014} then constructed a 1D steady-state model of a Kippenhahn-Schl\"uter-type prominence slab and demonstrated that under the action of gravity, a current layer was formed accompanied by a local velocity enhancement indicating the motion of material across the field lines.

To be more explicit, one of the cases explored by \citet{Low:2012b} considered a temperature variation across those flux surfaces being compressed by the heavy prominence material. These authors also considered for a moment that the compression may be significant enough for two adjacent field lines to be brought very close together in space. The perpendicular thermal conduction may therein also become important. However, the authors also note that the ratio $\epsilon$ between the perpendicular thermal conductivity $\kappa_\perp$ and resistivity $\eta$ has, of course, the limits of $\epsilon \ll 1$ and $1 \ll \epsilon$ corresponding to resistive or conductive behaviour, respectively. In our simulation the perpendicular thermal conductivity is assumed negligible and therein completely neglected, leading to $\epsilon=0$ everywhere. However, they also found that $\epsilon = 0.64 \times \mathrm{(plasma-)}\beta$ and thus provides a quick check for whether our assumption remains valid within the region between these compressed flux surfaces. The distribution of $\beta$ within the simulation domain is presented in the middle row of Figure~\ref{fig:rti_evol_low_res}. It is shown that the value of $\beta$ is of order 1 or less in the entire simulation domain, including the condensation region, and $\ll 1$ within the remainder of the evacuated flux rope. At the interface between the falling condensations and the lower, compressed field we find localised regions of $\beta \approx 0.1$. Hence, $\epsilon$ may be considered (much) less than one at the positions of these interfaces and hence resistive behaviour dominates over conductive. 


\begin{figure*}
	\centerline{\includegraphics[width=1\textwidth,clip=, trim=0 60 0 55]{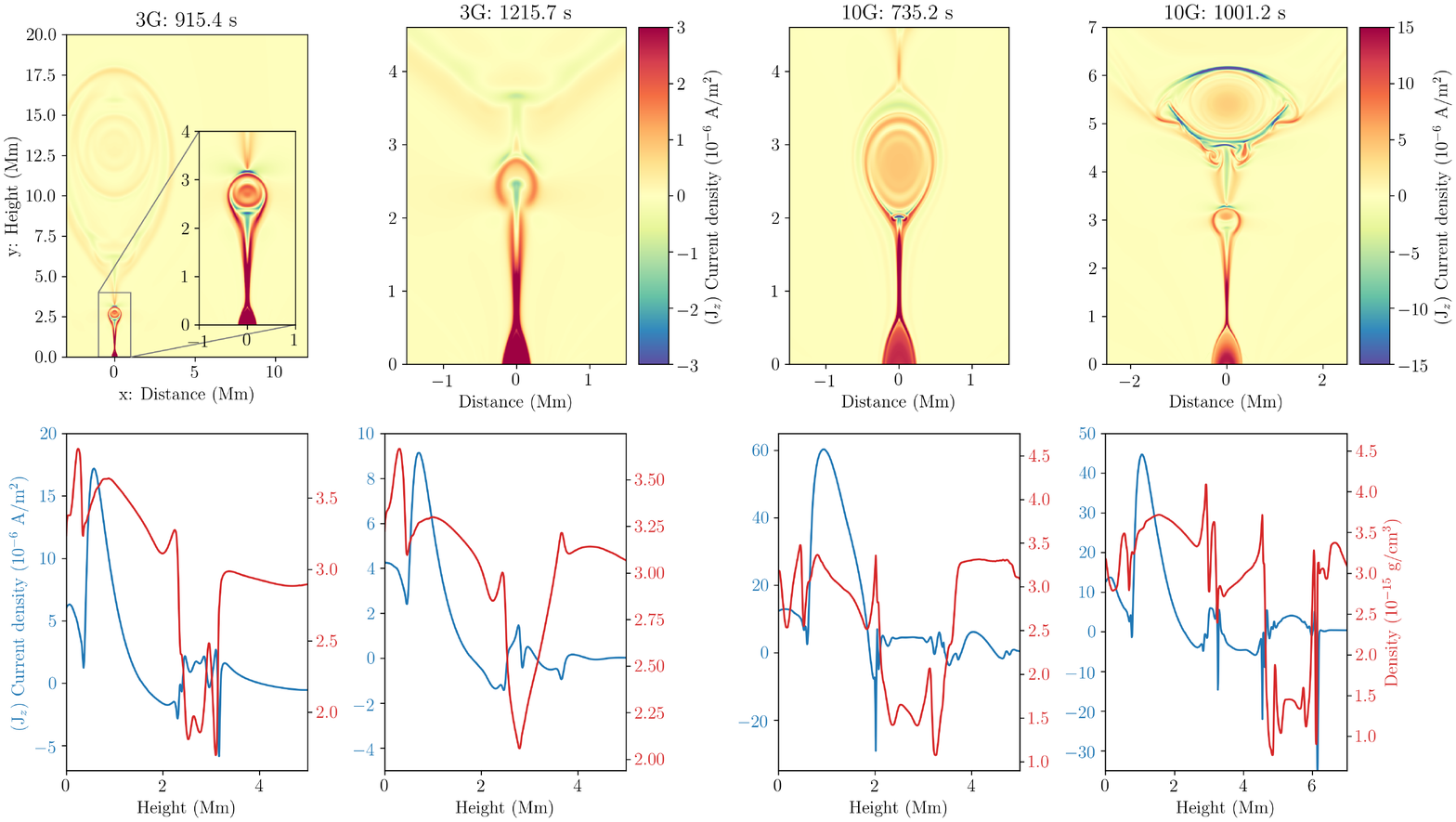}}
	\caption{The production of subordinate flux ropes below the main rope and their involvement with the supply of material to the prominence. The top panels describe the distribution of $J_\mathrm{z}$ current density. The bottom panels compare the current and plasma density distribution in a vertical cut through the current sheet at $x=0$. \textit{Left} \& \textit{Right}; The construction of two additional flux ropes ($t=915.4$, $1215.7$ \& $t=735.2$, $1001.2$~s) within the 3G and 10G cases, respectively}
	\label{fig:flux_feeding}
\end{figure*}

In the left panel of Figure~\ref{fig:low_2014_t} we show a snapshot at $t=8294.2$~s of the density and orientation of the magnetic field lines within the region of the falling material \citep[cf.][]{Low:2012a}. The panels on the right then detail the evolution ($7522.0<t<8294.2$~s) of the density, current density, vertical velocity, and temperature along a cut taken at $x=0$. The four cuts are presented as zoom-ins on the bow i.e., the leading edge, of the falling material, indicated by the increase/decrease in density/temperature by several orders of magnitude. At $t=7522.0$~s, the current density is enhanced at $\approx9$~Mm as the local field lines begin to be compressed ahead of the growing condensation. By $t=7779.4$~s we see that the vertical velocity of the falling monolithic structure has decreased and the current enhancement is now more localised to the bow of the falling material. As the simulation progresses further, a small and localised negative velocity enhancement is observed to grow just below the current enhancement.

Figure~\ref{fig:low_2014_3} then explores the same quantities as Figure~\ref{fig:low_2014_t} but for a later time $t=8654.56$~s. At this time, and as can be seen in the left panel of this figure, the density concentrations have formed three distinct fragments or isolated islands. The top panel then details the variation in the density, current density, vertical velocity, and temperature in a cut from 0~Mm to 8~Mm in height at $x=0$. The bottom three panels of Figure~\ref{fig:low_2014_3} focus on the leading edges of each distinct density concentration, the lowest referred to with a (1), the middle with a (2), and the highest with a (3).

\subsection{5.7~km resolution and a continous cooling curve}\label{ss:highres}
\subsubsection{Modified setup}
In this Section we revisit the formation and evolution of prominences by using a much higher numerical resolution, and for two different field strengths. To that end, we make full use of the \ac{AMRVAC} toolkit and adopt a base grid of 528~$\times$~528 with 3 additional levels of refinement permitting a maximum effective gridcell size of $\approx$~5.7~km. We also improve on our assumption of optically-thin radiative losses by exchanging the piecewise cooling curve of \citet{Hildner:1974} for the continuous solar metallicity cooling curve of \citet{Schure:2009}, wherein the lower temperatures are governed by the behaviour of H~{\sc i} \citep[see][and the \ac{AMRVAC} website\footnote{\href{http://amrvac.org/md_doc_radiative_cooling.html}{Radiative cooling in \ac{AMRVAC}}}]{Dalgarno:1972}. For the majority of the simulation time, the equations (\ref{eq:masscont})\,--\,(\ref{eq:induction}) are solved using the same numerical methods as described in Section~\ref{ss:num_met}. However, we also adopt a more-stable two-step, second-order \ac{TVDLF} method, with a Monotonized Central (or `woodward') limiter. We also artificially enforce an overall minimum temperature of 1000~K, far below the anticipated temperature of the prominence condensations to ensure, at all times, positive pressure. Similarly, we solve an auxiliary internal energy equation in tandem with the total energy equation such that we can guarantee positive pressures; in very low-$\beta$ plasmas the numerical error for total energy is of the order of the internal energy and hence the total energy may become negative. For $\beta<0.005$ we trust the solution to the internal energy, apply a linear combination of internal and total energy for $0.005<\beta<0.05$, and simply continue with total energy for $\beta>0.05$. The two surface magnetic field cases explored in this section adopt a $B_\mathrm{a}$ value of 3~G and 10~G, labelled `3G' and `10G' respectively. Finally, we choose to break the artificial left-right symmetry adopted in Section~\ref{ss:lowres} with the imposition of an arbitrary perturbation to both the 3G and 10G cases at $t=1945$~s of the form $v_\mathrm{x}=-v_\mathrm{max} \mathrm{sin}(\pi y/25)$, where $v_\mathrm{max}=12$~km~s$^{-1}$.


\subsubsection{Prominence formation and evolution}

In each case 3G and 10G, the formation of the flux rope follows the previous example wherein the linear, force-free coronal magnetic field initially relaxes before the bottom boundary drives their footpoints together. The initiation of reconnection at the point $x=0$ then builds the twisted magnetic field that is lifted above the surface as reconnection supplies flux to the forming rope. The increased resolution also allows us to resolve further reconnection-related fine-structure, formed within the current sheet underneath the main flux rope, and in both 3G and 10G cases, additional subordinate flux ropes are formed in this sheet.

The top panels of Figure~\ref{fig:flux_feeding} capture these ropes after formation and during their launch into the main flux rope, wherein they are later incorporated \citep[such dynamic mergers have previously been recorded in high-resolution numerical simulations \textit{e.g.},][]{Zhao:2017, Zhao:2019}. The particularly large dimensions of these ropes in comparison with previous studies is influenced by the fairly large $\eta$ value of 0.002 across all simulations within this work. As such, reconnection progresses much more easily than expected on the actual Sun. Nevertheless, this is analagous to the observations of `flux feeding' suggested to play a role in the mass-supply of prominences or in CME eruptions \citep[\textit{e.g.},][]{Liu:2012b,Zhang:2014,Zhang:2020}. The lower panels of Figure~\ref{fig:flux_feeding} detail the distribution of electric current density (into the plane \textit{i.e.}, $J_\mathrm{z}$) and mass density along a cut taken vertically at $x=0$.

The flux rope formation process provides a non-negligible upward acceleration to the flux rope and trapped plasma. Once the boundary driving and reconnection has ceased, the formed flux ropes begin oscillating vertically with a damped amplitude. For the 10G case, the oscillation has a period of $\approx~350$~s and persists for only a short while whereas the 3G case has a longer period of $\approx~900$~s and continues to oscillate for an extended period due to the relatively weaker magnetic field and restoring tension force therein. The symmetry is then broken in line with the aforementioned perturbation applied at $t=1945$~s. Shortly thereafter, the temperature within the core of the 3G flux rope begins to decrease rapidly and condensations form quickly ($t\approx2743$~s). For the 10G case, the larger magnetic field strength/current and associated Joule heating $\eta\vec{J}^2$ maintains the temperature of the plasma for an extended period of time until finally the temperature in one of the outer layers of the flux rope drops at $t\approx5162$~s, accompanied by the formation of condensations. In the following, we apply the same analysis as explored in Section~\ref{ss:lowres}.

\begin{figure*}
	\centerline{\includegraphics[width=1\textwidth,clip=, trim=65 199 20 180]{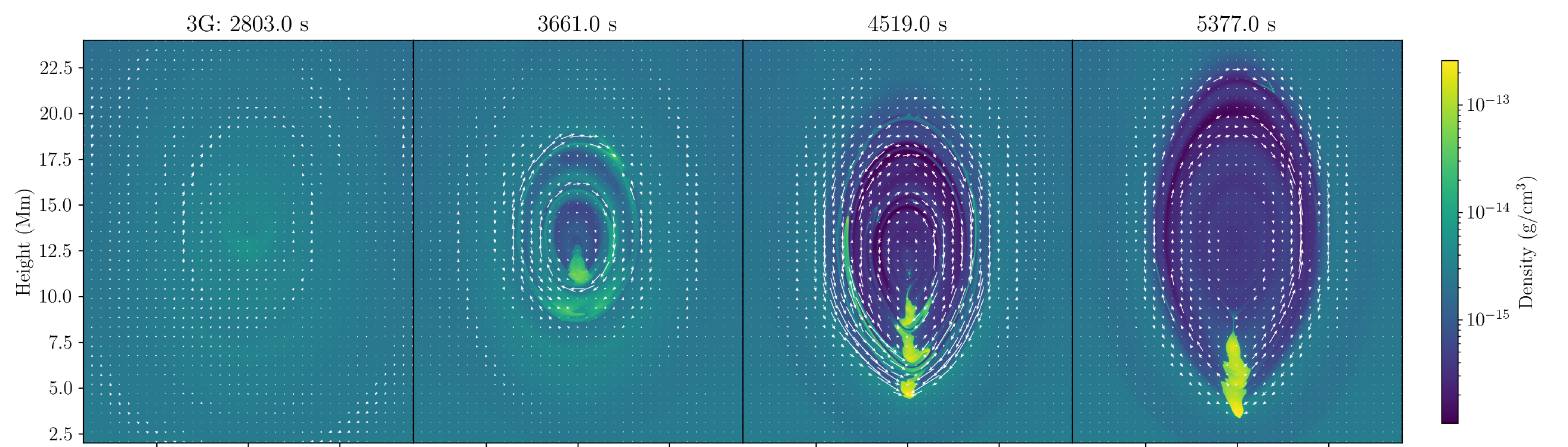}}
	\centerline{\includegraphics[width=1\textwidth,clip=, trim=65 170 20 196]{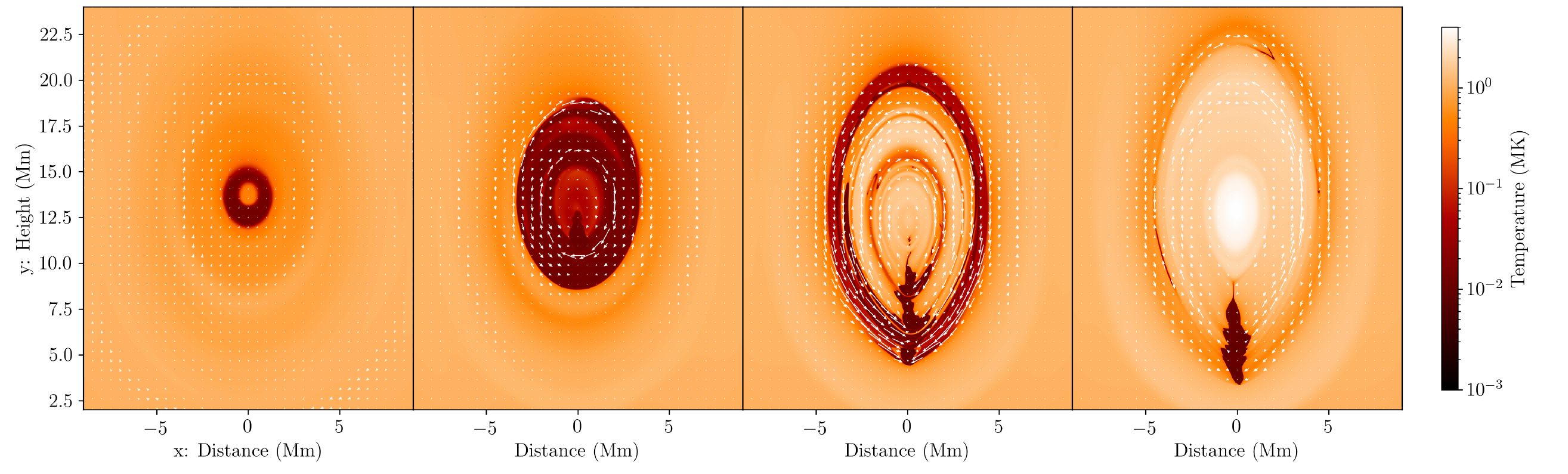}}
	\caption{Evolution of the condensing material within the 3G formed flux rope. The \textit{first} row details the evolution of the density, \textit{second}; temperature. Velocity quivers of relative magnitude are overplotted as before. A movie of this figure, including $N_{m, p}^{2}$ and baroclinicity maps, is available with the online version of this manuscript.}
	\label{fig:3G_condensation_evol}
\end{figure*}

\begin{figure*}
	\centerline{\includegraphics[width=1\textwidth,clip=, trim=0 160 0 155]{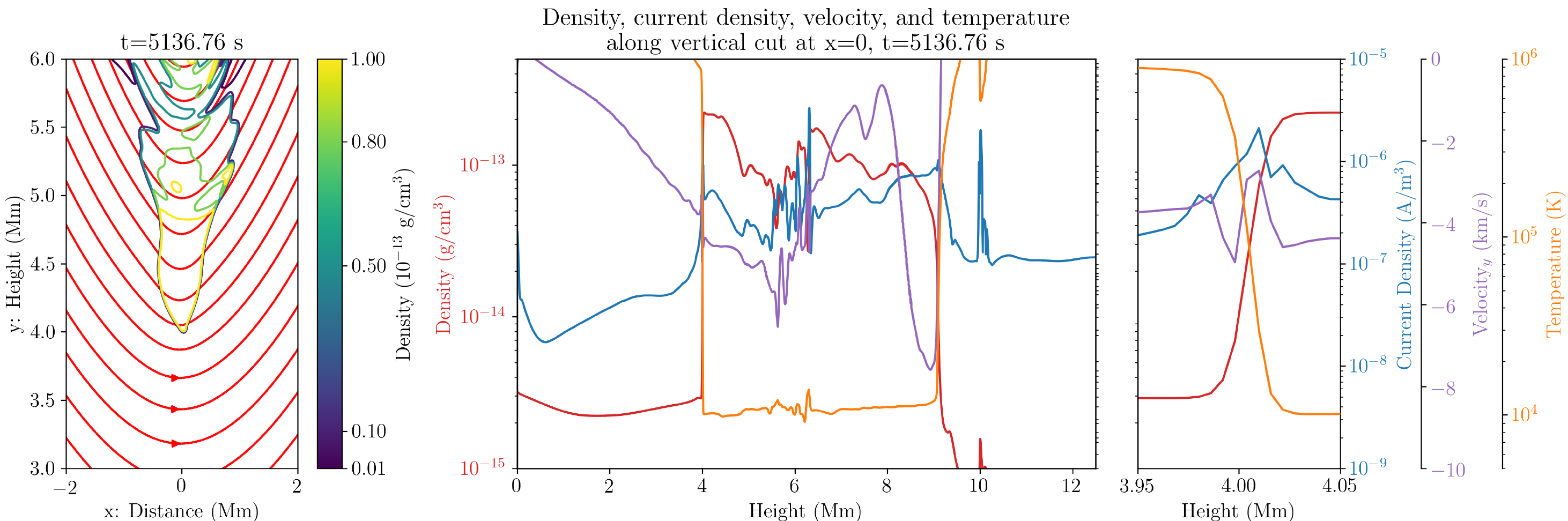}}
	\caption{Magnetic field and density map, and stratification of plasma density, current density, vertical velocity, and temperature within the 3G case along a cut taken vertically through $x=0$ at $t=5136.76$~s. \textit{Left}; a zoom in of the bottom of the prominence condensation. Magnetic field lines are plotted in red with density contours overlaid at thresholds according to the colorbar. \textit{Middle}; the stratification of the aforementioned plasma parameters between 0\,--\,12.5~Mm, isolating the entire prominence condensation. \textit{Right}; A zoom-in of the cusp of the condensation, showing the behaviour of the plasma parameters across the boundary.}
	\label{fig:low_2014_3G}
\end{figure*}

Figure~\ref{fig:3G_condensation_evol} presents the evolution of plasma density and temperature within the 3G case. The core of the flux rope is the first region to cool as it is the first region to be disconnected from the surface; thermal conduction is no longer transporting energy from the bottom boundary where the magnitude of the heating function $\mathcal{H}$ is largest. Indeed, the flux addition to the flux rope appears to be mimicked in the ordering of runaway cooling from the core to the outer-edges of the flux rope. Nearly identical to Figure~\ref{fig:lowres_TI}, we find that maps of $\partial \mathcal{Q}/\partial T \divisionsymbol \partial \mathcal{Q}/\partial \rho$ and $C$ for the 3G case indicate the decreasing temperature of plasma is driving the \ac{TI} from the core of the flux rope to its outer edges with time. At the time of initial condensation, the symmetry-breaking perturbation of the flux rope has not completely damped out and so the forming condensations carry some of this momentum. Nevertheless, the evolution of all cooled plasma is ubiquitously towards the concave-up and dipped portions of the flux rope topology. As in Section~\ref{ss:lowres}, we find velocities of material \textit{entering} the condensations as high as $\approx$~90~km~s$^{-1}$. Some packets of condensed plasma inherit significant momentum from the host rope. In combination with gravity these packets are observed to quickly slide to the bottom and are thrown back up to higher heights within the flux rope with velocities of the order 35~km~s$^{-1}$ as they pass the topological dips. The condensing material displays a variety of `sloshing' and oscillatory motions as its kinetic energy is damped and eventually the material comes to near-rest at the bottom of the rope. Interestingly, throughout the evolution of the system we see material condensing at locations throughout the flux rope \textit{i.e.}, not just as a continuous process but rather as isolated condensations even at higher heights \citep[see, for example, the $t=5377$~s panel and the localised temperature decreases at $h\approx10$~\&~$21$~Mm and compare with][]{Kaneko:2015}.

At $t=3661~s$ we can see that the condensing material initially cools the rope, almost in its entirety, down to an average temperature of between 10,000\,--\,100,000~K. Once the material begins to collect in the concave-up dips of the magnetic flux rope ($t=4519$~s), we see that the background temperature of the rope returns to $\approx~1$~MK. At the later time of $t=5377$~s, the core of the flux rope has begun to heat to $\approx~4$~MK. Towards the end of the condensation process, unlike as observed with the configuration in Section~\ref{ss:lowres}, the whole prominence begins to fall with an average velocity of $\approx$~2~km~s$^{-1}$, likely due to the cross-field slippage and reconnection in its wake. It is this reconnection that is responsible for heating the core of the 3G flux rope seen in the snapshot at $t=5377$~s. The peak magnetic field strength is recorded within the condensed plasma portion of the 3G flux rope with a value of $\approx~4.1$~G.

Comparing the distribution of the $N_{m, p}^{2}<0$ locations with regions of high baroclinicity, the $N_{m, p}^{2}<0$ solutions appear to be distributed throughout both the condensing region and the surrounding flux rope whereas the baroclinitity signatures appear localised to the condensing region itself (see the online movie associated with Figure~\ref{fig:3G_condensation_evol}). At later times, the $N_{m, p}^{2}<0$ signatures remain largely spread out whereas the baroclinicity has become localised once more to the positions of the prominence condensations.

\begin{figure*}
	\centerline{\includegraphics[width=1\textwidth,clip=, trim=65 201 20 190]{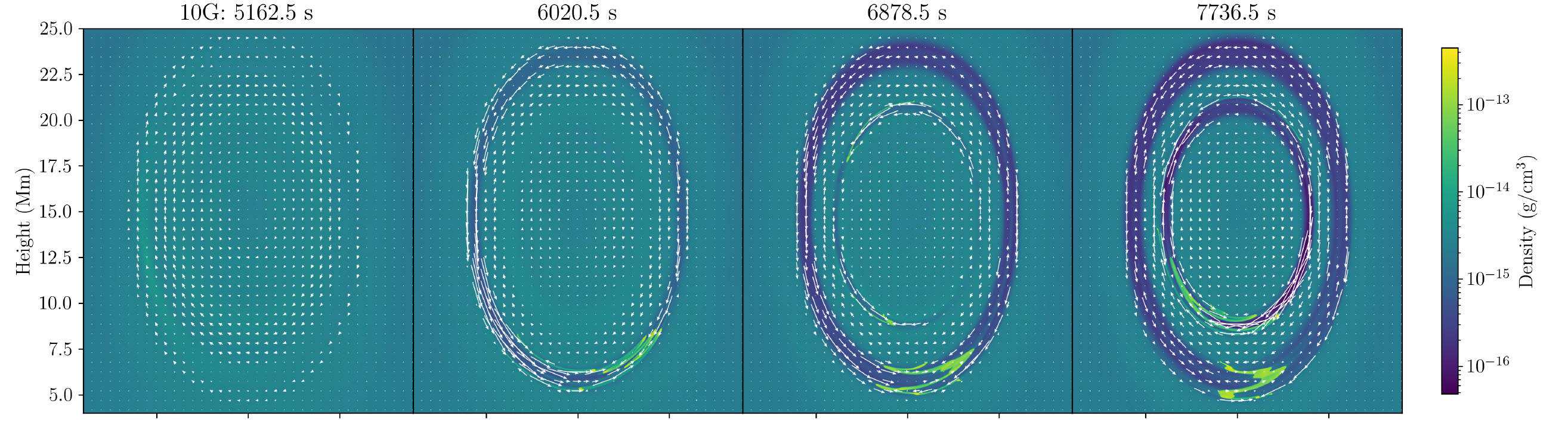}}
	\centerline{\includegraphics[width=1\textwidth,clip=, trim=65 170 20 201]{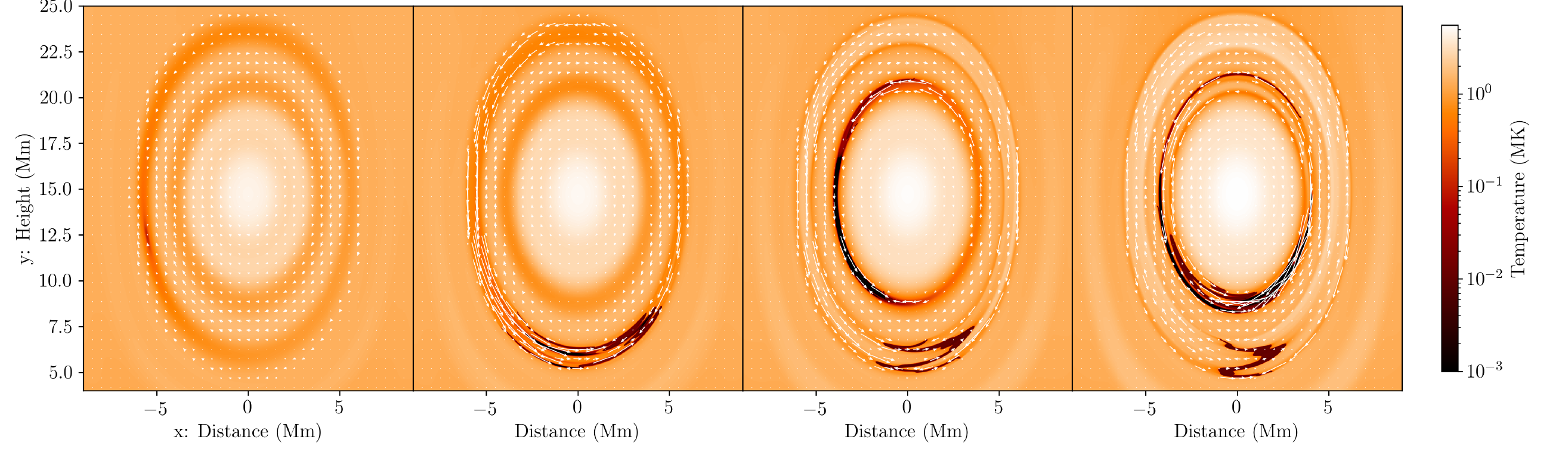}}
	\caption{Evolution of the condensing material within the 10G formed flux rope. The \textit{first} row details the evolution of the density, \textit{second}; temperature. Velocity quivers of relative magnitude are overplotted as before. A movie of this figure, including $N_{m, p}^{2}$ and baroclinicity maps, is available with the online version of this manuscript.}
	\label{fig:10G_condensation_evol}
\end{figure*}

\begin{figure}
	\centerline{\includegraphics[width=0.5\textwidth,clip=, trim=120 0 300 0]{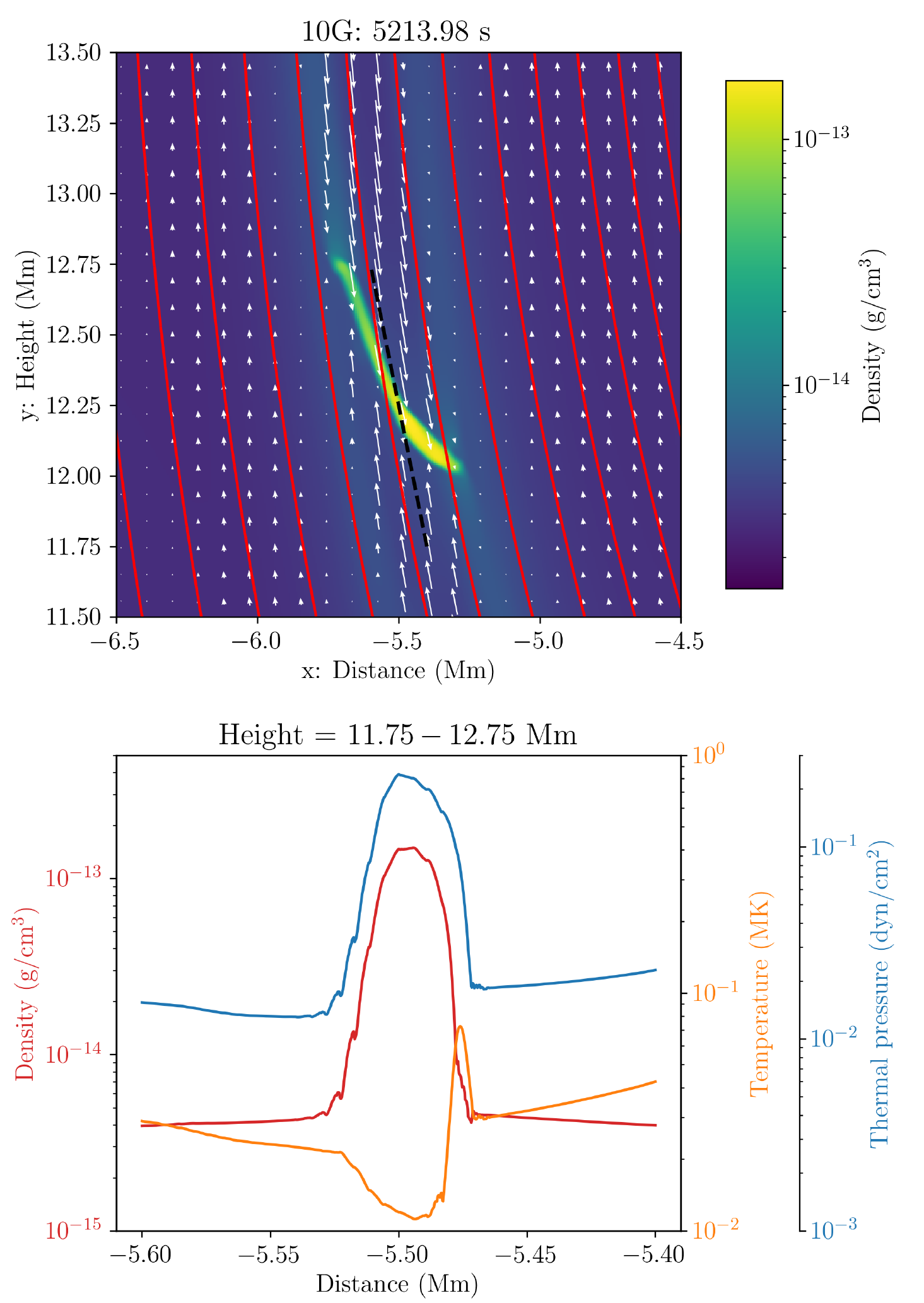}}
	\caption{Density, temperature, and pressure cross-section of the initial filamentary condensation within the 10G prominence formation case. \textit{Top panel}; the distribution of density within the forming filamentary condensation. Velocity quivers of relative magnitude are overlaid in white. Magnetic field lines are overlaid as solid-red lines. \textit{Bottom panel}; the density, temperature, and pressure profiles across the condensation cross-section, taken along the dashed-black line in the top panel. Profiles computed using yt \citep[yt-project:][]{Turk:2011}.}
	\label{fig:10G_condensation_dens_zoom}
\end{figure}

Towards the end of the simulation, the condensations of the 3G case coalesce at the bottom of the magnetic topology and form once more (cf. the lower resolution case) a monolithic structure. Recall that we chose to break the symmetry by applying a perturbation. At the time of the formation of the monolithic structure (see the $t=5377.0$~s snapshot of Figure~\ref{fig:3G_condensation_evol}), residual yet small-scale oscillations persist, requiring delicate consideration when applying the same analysis as presented in Figures~\ref{fig:low_2014_t}~\&~\ref{fig:low_2014_3}. Figure~\ref{fig:low_2014_3G} presents the cut of plasma density, current density, vertical velocity, and temperature along a vertical cut at $x=0$ at a time when the bulk oscillations in both the vertical and horizontal directions are minimal. The complete monolithic structure is captured at this time, clearly indicated in the middle panel with the increase (decrease) in density (temperature) between 4\,--\,9~Mm. The right panel then focusses on the bottom 3.95\,--\,4.05~Mm of the condensation wherein the field lines are compressed as a result of the plasma weight. Note here, that the compression appears to be to a lesser degree than the case explored in Section~\ref{ss:lowres}, nevertheless a clear current enhancement is also present at the cusp in addition to the local velocity enhancement in the negative vertical direction. The increase in the resolution of the simulation in the 3G case has enabled the local enhancements in both current and vertical velocity to be better resolved than in Figures~\ref{fig:low_2014_t} or \ref{fig:low_2014_3}.

Figure~\ref{fig:10G_condensation_evol} presents the evolution of plasma density and temperature within the 10G case. The core of the 10G flux rope is significantly hotter than the 3G case (up to 5~MK), owed to the larger magnetic field driving enhanced Joule heating. The large currents involved in the formation process \textit{i.e.}, the subordinate flux ropes presented in Figure~\ref{fig:flux_feeding}, provide significant current to a shell around the core. Hence, both of these regions remain hotter than the corresponding regions in the 3G case. This has the primary consequence of preventing these regions from cooling and hence condensations do not begin in the core but instead in one of the non-Joule-heated outer layers. The structure and shape of the initial condensations within the 3G case appear to be more global before becoming localised, whereas the initial condensation in the 10G case begins far more localised. The formation of the condensation starts to collect material along a flux surface with a velocity of order 70~km~s$^{-1}$, before gravity accelerates it to an average velocity of 25~km~s$^{-1}$ towards lower heights. Most strikingly, the orientation of this initial filamentary condensation is oriented perpendicular to the host magnetic field, a zoom-in is presented in the top panel of Figure~\ref{fig:10G_condensation_dens_zoom}. At $t=6878.5$~s, we see that two additional condensations are in the process of falling to the concave-up portions of the magnetic topology after having formed at two very different heights. The first and lowest of these two additional condensations forms at a similar height to the first condensation at $t=5162$~s, whereas the second occurs at a far more elevated height of $\approx~20$~Mm, before they both fall to the bottom of their flux surface. Both of these condensations have a contraction velocity of order 90~km~s$^{-1}$ \citep[driven by varations in (thermal/ram) pressure cf. ][]{Claes:2020}. The lower of the two condensations falls with a maximum velocity of $\approx$~20~km~s$^{-1}$. Interestingly, the initial contraction and subsequent fall of this lower condensation leaves a very low pressure region in its wake. The higher condensation now falling under gravity is also accelerated by this low pressure void, reaching velocities as high as 70\,--\,80~km~s$^{-1}$. As was found for the 3G case the peak magnetic field strength is recorded within the condensed plasma portion of the 10G flux rope, with a value of $\approx~10.7$~G.

\begin{figure*}
	\centerline{\includegraphics[width=0.4\textwidth,clip=, trim=110 60 45 60]{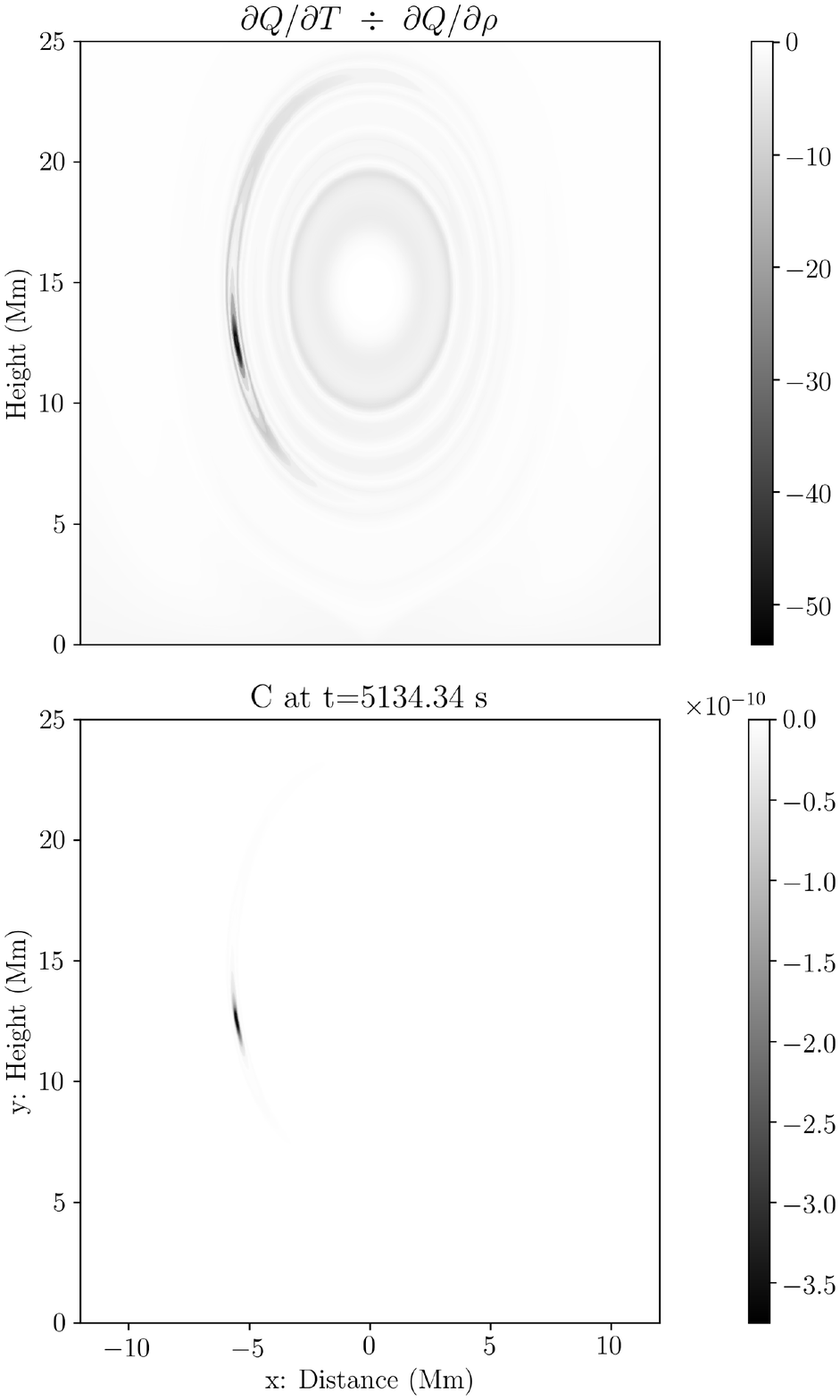}
		\includegraphics[width=0.4\textwidth,clip=, trim=110 60 45 60]{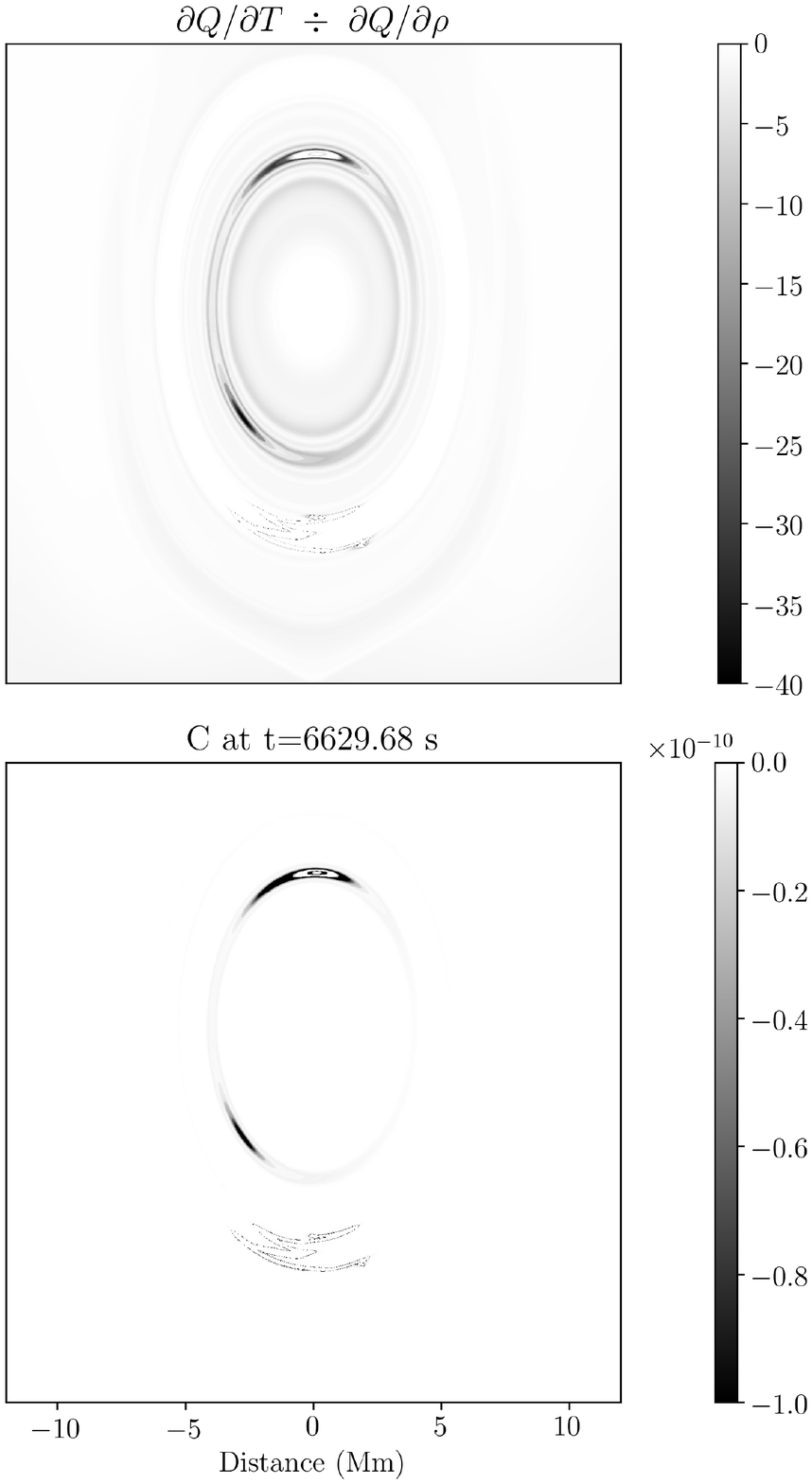}}
	\caption{Locations of thermal instability within the 10G prominence just prior to the two main episodes of condensation formation (\textit{Left}; 5134.34~s and \textit{Right}; 6629.68~s). \textit{Top}; Ratio of thermal to density influence on the varying energy loss $\mathcal{Q}$ term in eq.~(\ref{eq:econt}). \textit{Bottom}; Spatial map of the isochoric linear stability metric $C$ (eq.~\ref{eq:C}) for the thermal mode instability.}
	\label{fig:10G_TI}
\end{figure*}

Figure~\ref{fig:10G_TI} presents the maps of $\partial \mathcal{Q}/\partial T \divisionsymbol \partial \mathcal{Q}/\partial \rho$ and $C$ (wherein $k$=0.1~Mm, cf. Figure~\ref{fig:10G_condensation_dens_zoom}) for the 10G case at times just prior to the two episodes of condensation formation ($t=5134.34$~\&~$6629.68$~s). The maps of the \ac{TI}, constructed as maps of the isochoric $C$ metric, elegantly isolate the exact locations (both position and extent) of the forming condensations. Furthermore, and consistent with the other cases explored in the manuscript, we find that the early stage of the \ac{TI} within the condensing plasma of the 10G case is driven by the temperature variations. In fact, the resulting filamentary condensations of order $\rho=10^{-13}$~g/cm$^{3}$, $T=10,000$~K, $p=0.2$~dyn/cm$^{2}$ are directly comparable to the density structures formed by MHD slow-wave induced thermal instability modes \citep[this will be discussed in the Section~\ref{s:discussion} but see][]{Claes:2019, Claes:2020}.

Finally, inspecting the location of $N_{m, p}^{2}<0$ and large baroclinicity values (and variations therein), one finds satisfactory agreement with the locations of the prominence condensations (see the online movie associated with Figure~\ref{fig:10G_condensation_evol}).

\subsubsection{Visualising small-scale evolutions}

The increased resolution employed in the 3G and 10G cases lead to complex non-linear evolutions. Figures~\ref{fig:3G_condensation_evol}~\&~\ref{fig:10G_condensation_evol} are accompanied by movies in the online version of this manuscript. We include some \ac{halpha} visualisations of the evolution following the methods of \citet{Heinzel:2015} \citep[see the explicit implementation in the appendix of][]{Claes:2020}. For all visualisations, the prominence material is synthesised using the 10~Mm table of Table~1 in \citet{Heinzel:2015}.

\begin{figure*}
	\centerline{\includegraphics[width=1\textwidth,clip=, trim=0 70 0 70]{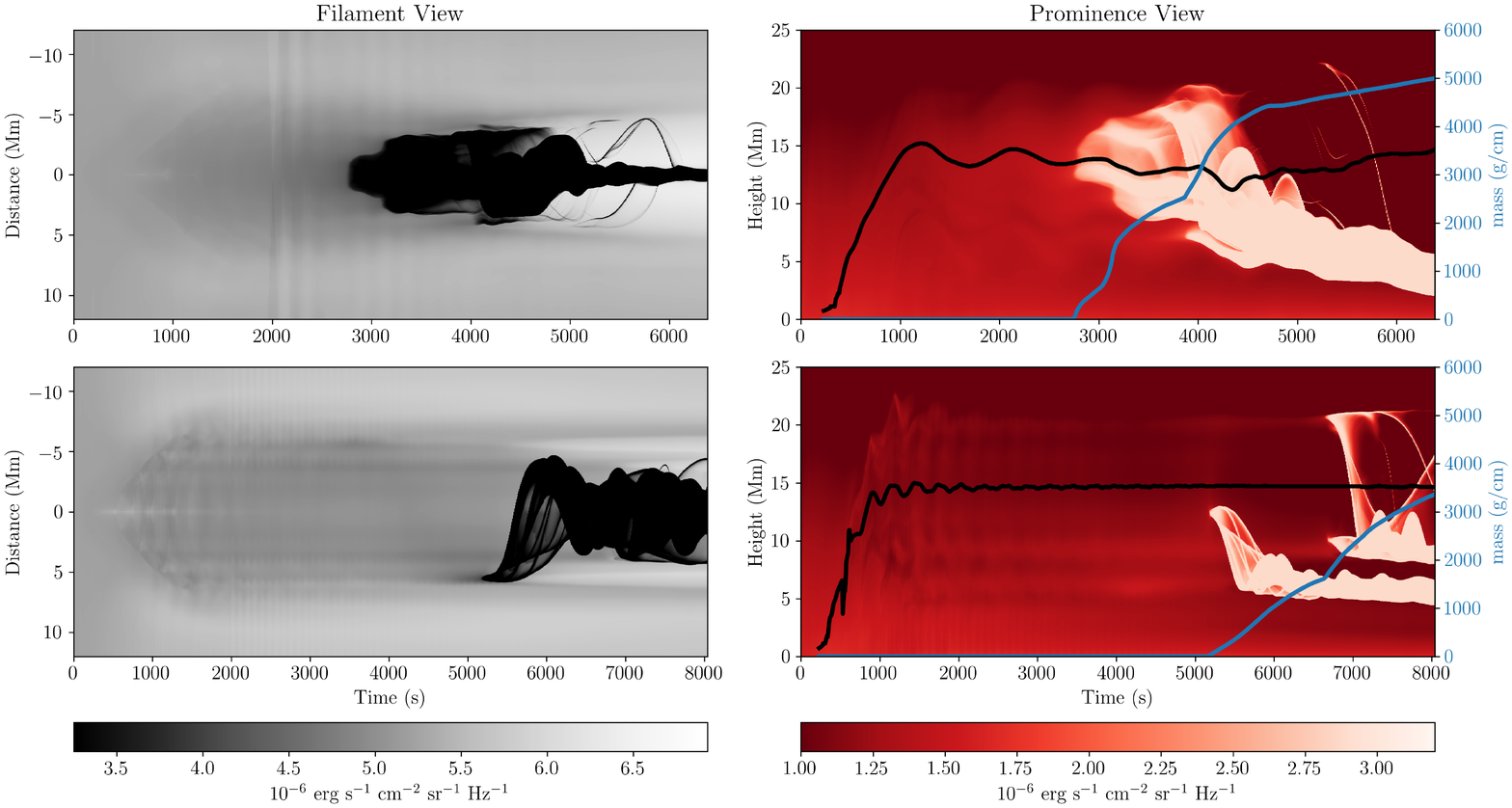}}
	\caption{\textit{Left}; Filament and \textit{Right}; prominence \ac{halpha} visualisations of the 3G (top) and 10G (bottom) cases using the method outlined in \citet{Heinzel:2015}. \textit{Right}; The evolution in the flux rope O-point and prominence mass are overplotted on the prominence evolution as the solid black and blue lines respectively. The large jump in altitude experienced by the 10G case is caused by the injection of the larger of two subordinate ropes into the main forming rope (cf. Figure~\ref{fig:flux_feeding}).}
	\label{fig:visualisation}
\end{figure*}

Figure~\ref{fig:visualisation} presents the \ac{halpha} appearance of the forming and evolving 3G and 10G cases from both a top-down (filament) and side-on (prominence) perspective. The flux rope formation period $0<t<1516$~s, associated oscillations, induced perturbation $t=1945$~s, and their damping can be visually identified as intensity variations within each visualisation. During the formation phase of the 10G case, the corresponding filament displays longitudinal oscillations across the width of the flux rope indicated by the periodic broadening of its cross section between $1000<t<1800$~s. The transverse, horizontal, and persistent oscillations induced by the symmetry-breaking perturbation are well captured within the filament visualisation of the 3G case. The transverse, vertically oriented oscillations induced by this formation process are present within the prominence visualisations of either case. To emphasise these vertical oscillations and their relation to the forming condensations, the position of each flux rope's axis (hereafter `O-point') are overplotted on the prominence visualisations. In each case, the formation and rise of the flux rope is captured along with the resulting vertical oscillations that are weakly damped in the 3G case, or strongly so in the 10G case. The sudden jump in the height of the 10G O-point is a result of the injection of the larger of two subordinate ropes presented in Figure~\ref{fig:flux_feeding}. In addition, we have overplotted the time evolution of total mass per unit length within the forming prominences, defined as the mass with a temperature below $T=10^5$~K.

\section{Discussion}\label{s:discussion}
\subsection{What drives the evolution of prominences in the early stages of their formation?}

This manuscript focuses on the mechanisms for, and drivers of, the formation and evolution of prominences within the solar atmosphere through numerical simulations. The host magnetic flux rope is constructed in-line with the assumption of photospheric diffusion and magnetic cancellation at the intersection of two separate flux regions of opposite polarity \citep[cf.][]{Martin:1985,vanBallegooijen:1989,vanDriel:2003}. The footpoints are converged and sheared by imposing a velocity on the lower boundary following \citet{Kaneko:2015} to investigate the levitation-condensation process. The result, as detailed in Figure~\ref{fig:den_evol_low_res}, is a flux rope with a left-handed twist that is progressively elevated through the solar atmosphere as a consequence of continued reconnection. In all cases explored here, the formation process leads to a vertical oscillation of the constructed flux ropes, as recorded in Figure~\ref{fig:visualisation}. The 31~km gridcell and `3G' cases experience a prolonged period of transverse oscillations with larger amplitude as their magnetic field is relatively weak in comparison with the `10G' case that displays much smaller amplitude oscillations that are quickly damped.

The first case explored in Section~\ref{ss:lowres} was the 31~km grid step study under symmetric conditions and assuming the piecewise cooling curve of \citet{Hildner:1974}. The resulting condensations appear symmetrically on either side of $x=0$ and with a velocity vector oriented towards the concave-up dips of the magnetic topology. In line with the conclusions of \citet{Kippenhahn:1957}, the cooling plasma appears to seek equilibrium where it can be supported against gravity by the tension of the magnetic field. The appearance of fine structure at this time, which was not reported in the study of \citet{Kaneko:2015}, is likely due to our more realistic handling of the low temperature limit. Indeed, they assumed a floor temperature of 0.3MK, far above the temperatures measured within prominences and more indicative of the cool yet diffuse plasma of the lower corona. The gradual evolution of this condensing, finely-structured plasma towards the \citet{Kippenhahn:1957} equilibrium is actually influenced by a plethora of plasma instabilities, the interesting question is then of course which is (are) the most relevant? 

Figures~\ref{fig:condensation_evol_low_res}, \ref{fig:lowres_TI}, \ref{fig:rti_evol_low_res}, \ref{fig:low_2014_t}, \& \ref{fig:low_2014_3} focus on the physical processes driving the evolution. According to Figure~\ref{fig:condensation_evol_low_res}, the formation of the prominence is the end result of various interacting physical processes. Once the flux rope has formed, the density of the plasma within the flux rope can be considered a flux function \textit{i.e.}, constant along a given flux surface but varied with distance from the flux rope axis. \textit{Small} density variations along a given flux surface caused by \textit{e.g.}, perturbations are then said to be \ac{CCI}(-like) unstable, causing a redistribution of plasma to restore equilbrium. This is well-captured in the maps of $N_{m, p}^{2}<0$ prior to the appearance of discrete condensations. The maps of isochoric $C$ presented in Figure~\ref{fig:lowres_TI} indicate the plasma in the core of the flux rope then undergoes the \acf{TI}, driven by the temperature in this region decreasing due to `runaway' radiative losses \citep[see Figure~\ref{fig:rad_evol} and cf.][]{Parker:1953, Field:1965, Klimchuk:2019b, Antolin:2020}. The subsequent formation of dense and cool regions additionally enforces the \ac{TI} due to the $\rho^2$ dependence of $\rho\mathcal{L}$, but the now-large variations of plasma density along a given flux surface halts any valid comparison to the \ac{CCI}. Nevertheless, the $N_{m, p}^{2}<0$ metric remains valid for evaluating regions within the flux rope that are evolving as a consequence of being out of pressure balance. As expected, the field-projected pressure scale height responsible for enforcing hydrostatic equilibrium along a flux surface appears to shrink as the material dramatically cools (leading to the appearance of $N_{m, p}^{2}<0$ regions within the flux rope as seen in Figure~\ref{fig:3G_condensation_evol}). This out-of-equilibrium plasma then falls along field lines in the direction of the shrinking scale height \textit{i.e.}, towards the concave-up portions of the magnetic topology. At time $t=7264.6$~s, the material within the outermost regions of the condensing region (compare with the distribution of optically-thin radiative losses in Figure~\ref{fig:rad_evol}) appears to undergo rotational motions as expected to occur when local baroclinicity is analysed. As the simulation evolves, this baroclinicity is then increasingly concentrated at the underside of the monolithic structure. The same may be said for the regions of $N_{m, p}^{2}<0$, however additional regions remain present elsewhere and at higher heights within the condensing region. Therefore, it appears that within this first setup: the \ac{CCI} is largely responsible for the evolution of plasma prior to the appearance of condensations, the \ac{TI} drives the stable formation of condensations, and these condensations fall to the concave-up portions of the magnetic topology as a consequence of being out of pressure balance on a given flux surface. The enhanced signatures of baroclinicity located within the condensing material then indicates that the material evolving due to being out of pressure balance is also rotating. Although baroclinic signatures are expected in the wake of material evolving under the \ac{RTI}, a direct comparison is not possible here due to the orientation of the 2.5D domain with respect to the magnetic field vector. However, the explicit coincidence between baroclinic signatures and the condensed material indicates such rotations may be identifiable within particularly-high resolution observations of prominences. The final result, as shown in the $t=7607.8$~s snapshot of Figure~\ref{fig:condensation_evol_low_res}, is the formation of a monolothic prominence embedded within a density-depleted and cold coronal cavity.




Following the formation of the large-scale, coherent prominence, discrete blobs of plasma \textit{i.e.}, `plasmoids' are observed to fall from its underside, eventually developing pinchpoints in their wake and splitting off entirely after several episodes of reconnection \citep[see the top panels of Figure~\ref{fig:rti_evol_low_res} and cf.][]{Liu:2012a}. In order for such dynamics to occur, the gravitational forces have to exceed the magnetic tension forces so as to sufficiently drag the host field lines towards the bottom of the simulation domain. Indeed, we see from the second row of Figure~\ref{fig:rti_evol_low_res} that the values of $\beta$, taken as a first order approximation to this effect, within the falling plasmoids were of the order, or greater, than one. The bottom row of the same figure then details the distribution of baroclinicity wherein the negative-left and positive-right signatures at the edges of the monolithic structure and falling plasmoids can be attributed to the shape of the general pressure gradient within the flux rope. Then, for the falling plasmoids, theory suggests clockwise (positive) and anticlockwise (negative) rotations are expected to be present on the left/right of structures falling and evolving under the influence of gravity, respectively. Indeed, smaller structures can be identified at the very periphery of the falling blobs. This suggests that although this 2.5D simulation does not contain distinct signs of the large-scale `roll-up' of plasma involved in the \ac{RTI}, there exists rotational motions at the edges of these falling plasmoids \citep[cf.][]{Hillier:2012b}. 

Figures~\ref{fig:low_2014_t}~\&~\ref{fig:low_2014_3} focus on the distribution and evolution of plasma and magnetic parameters within these plasmoids in relation to the theory proposed by \citet{Low:2012b}. At all times explored in Figures~\ref{fig:low_2014_t}~\&~\ref{fig:low_2014_3}, the leading edge of the monolith and Blob (1), respectively, display a current enhancement consistent with the compression of field lines ahead of them. As this current enhancement grows, we see the localisation of a negative velocity enhancement that appears to bridge those cold flux surfaces containing the high-density prominence plasma with those below that are both hot and of low density. By contrast, the 1D study of this process by \citet{Low:2014} predicted the current enhancement would be co-spatial with the velocity enhancement, whereas here there is a clear displacement between the two signatures, with the velocity signature always leading the current signature. Blobs (2) and (3) then contain the same current enhancement but the velocity signature is significantly positive in comparison. Blobs (2) and (3) each exist above a reconnection point and in the wake of the blob below it. As such, it is conceivable that the reconnection process would provide a non-negligible positive velocity component between each blob as the field associated with the reconnection relaxes upwards. Hence, the cross-field diffusion described by \citet{Low:2012a,Low:2012b}, if occuring for Blobs (2) and (3), may be supressed in these regions or simply be of a lengthscale below the resolution of the numerical grid ($<62$~km) in this case. Crucially, however, the magnitude of the negative velocity measured ahead of the falling monolithic structure and Blob (1) is of the order and less than the velocity of each structure itself. 


To further explore the initial results and conclusions of Section~\ref{ss:lowres}, in Section~\ref{ss:highres} we moved to consider two higher-resolution (5.7~km resolution), continuous cooling curve, symmetry-broken simulations of prominence formation and evolution proccesses above either a 3 or 10~G photospheric magnetic field. The general evolution of these 3G and 10G cases are presented in Figures~\ref{fig:3G_condensation_evol}~\&~\ref{fig:10G_condensation_evol}, respectively, with \ac{halpha} visualisations presented in Figure~\ref{fig:visualisation}.

The initial formation process for the 3G case is driven in the same way as the lower resolution case from Section~\ref{ss:lowres}, and indeed the resulting evolution is nearly identical with the exception of the `flux-feeding' structures of Figure~\ref{fig:flux_feeding}. The global flux rope structure also behaves in much the same way, experiencing long period, weakly damped vertical and additional horizontal oscillations in response to both the formation and the symmetry-breaking perturbations \citep[see Figure~\ref{fig:visualisation} and cf. ][]{Molowny-Horas:1999, Terradas:2002}. Furthermore, the catastrophic cooling within the 3G case is also concentrated first at the core of the flux rope before spreading out to the outer edges with cool condensations forming in its wake. As such, these condensations initially form at almost all locations within the rope before falling to the concave-up portions of the magnetic topology. It is important to note here that the combination of the realistic cooling curve \citep[][]{Schure:2009} and the symmetry-breaking perturbation lead to the formation of these condensations at a much earlier time than the lower resolution, \citet{Hildner:1974} cooling curve, and symmetric setup.

The evolution of the prominence in the 10G case differs significantly from the two previous cases involving a 3~G photospheric field. First and foremost, the increase in magnetic field leads to an order of magnitude increase in the magnetic energy within the simulation. As such, the formation of the host flux rope occurs faster, and the associated oscillations induced by this process have both a much smaller amplitude and are damped much faster, see the O-point evolution in the bottom-right panel of Figure~\ref{fig:visualisation}. Then, by direct comparison with the 3G case, the condensations appear delayed in the 10G case. Indeed, thermal conduction is more efficient along those field lines of higher magnitude and as such any temperature perturbations that may therein lead to condensations are more-easily stabilised. In the absence of any additional physical drivers of these condensations, a significant portion of an entire flux surface would be required to cool \textit{enough} for catastrophic runaway cooling to initiate locally. Despite this, the size and shape of the initial condensations in the 10G case are far smaller and more-localised than for the 3G case as shown in the \ac{TI} maps of Figure~\ref{fig:10G_TI}.

Upon the eventual formation of cool condensations within the 10G prominence, the characteristic size and shape of the initial condensations are far smaller and more-localised than those found in the 3G case. Figure~\ref{fig:10G_condensation_dens_zoom} presents a zoomed-in view of the first condensation to form in the 10G case, in which it is immediately clear that multiple flux surfaces are involved across the length of this highly-localised condensation. Indeed, this figure facilitates a direct comparison with those filamentary condensations previously found in the studies by \citet{Claes:2019,Claes:2020}. Although these authors adopted a highly-idealised setup, their work demonstrated that the selective excitation of the thermal \ac{MHD} mode results in the formation of these structures. Furthermore and more generally, the authors suggested that these features may be formed as a natural consequence of radiative cooling + any perturbation \textit{e.g.}, the perturbation experienced by the entire flux rope shortly after formation, or the one applied to deliberately break the symmetry at $t=1945$~s. Of course, the real Sun is highly dynamic and perturbations from \ac{MHD} waves occur ubiquitously, therein suggesting the ubiquitous formation of thermally-driven condensations. 

We find that the \ac{CCI}, \ac{TI}, $N_{m, p}^{2}<0$, and baroclinicity metrics, when applied to the 3G and 10G cases, indicate the identical relationship between the individual metrics and the initial stages of prominences' evolution as was found in the 31~km gridstep setup. The \ac{CCI} maps regions within the flux ropes wherein plasma is evolving slowly in response to changing pressure gradients induced by the global oscillations traced in Figure~\ref{fig:visualisation}. The isochoric $C$ value of the \ac{TI} is negative and large in locations where condensations are in the process of forming, particularly striking examples are presented in Figure~\ref{fig:10G_TI} for the highly localised condensations of the 10G case. After the appearance of distinct and discrete condensations, the flux ropes become filled with localised regions of $N_{m, p}^{2}<0$ indicating plasma evolving throughout the flux rope due to varying temperature and associated pressure gradients. The condensations themselves are then cospatial with strong baroclinic signatures revealing that even those condensations on the smallest scales are rotating in tandem with evolving towards the concave-up dips of the magnetic topology.

As time progresses for the higher resolution setups, the locations of the $N_{m, p}^{2}<0$ plasma become more-evenly distributed, consistent with the observation of plasma cooling throughout the flux rope and falling towards the concave-up dips of the magnetic topology. Once material fully condenses to the concave-up dips we see it no longer highlighted as a region of $N_{m, p}^{2}<0$. The material within the prominence has once more regained pressure equilibrium \citep[][]{Blokland:2011a,Blokland:2011b}. This material does, however, appear to continue to undergo rotations once within the prominence as indicated by the persistent concentrations of strong baroclinicity at the edges of the formed prominence. On the other hand, the localisation of the signatures to the periphery of the prominence could indicate a purely local rotation induced by the ongoing condensation of material onto the already-condensed prominence.

Finally, Figure~\ref{fig:low_2014_3G} presented the profile of plasma density, current density, vertical velocity, and temperature along a vertical cut along $x=0$ at a time within the 3G case when the oscillation of the formed monolithic structure was minimal. Previously, it was noted that the compression of the field lines is much smaller than in the previous case of Section~\ref{ss:lowres}, nevertheless the signature consistent with cross-field mass diffusion is still found \citep[][]{Low:2012a,Low:2012b}. Specifically, a local current and negative velocity enhancement is present at the leading edge of the falling monolith as predicted in the 1D steady-state study of \citet{Low:2014}. Once again, in comparison with the 1D steady-state model of \citet{Low:2014}, we see that the velocity and current enhancements are not cospatial; the velocity enhancement is located below the current enhancement as was the case in the lower resolution simulation. By definition, however, the authors state that the amount of material diffused from one flux surface to another in this way is quantitatively finite but as-yet unbounded, and so an explicit comparison is both non-trivial and outside the scope of this study. These results are, nevertheless, consistent with the hypothesis of \citet{Low:2012a,Low:2012b}~\&~\citet{Low:2014} and indicate that this physical process should be studied in more detail.

\subsection{Additional features and dynamics}
The primary focus of this manuscript was in the physical processes involved in the formation and evolution of prominence condensations formed by catastrophic radiative losses, however these simulations contain a number of additional and interesting features that we will briefly summarise here.

The lower-resolution case explored in Section~\ref{ss:lowres} may be directly compared to the previous study of \citet{Kaneko:2015}, wherein the main differences are (1) a different treatment of the upper boundary (K\&Y: `free' boundary; here: fixed density, pressure, and velocity + `extrapolation' of magnetic quantities); (2) a choice of background heating function that resembles the assumed photospheric source for coronal heating \citep[cf.][]{Fang:2013}, and (3) the relaxation of their limit on a minimum temperature to below 0.3~MK. The difference in the choice of the upper boundary is unlikely to have a large impact on the resulting dynamics within the forming condensations. The different heating functions, on the other hand, are able to explain the delay in the formation encountered in the case of Section~\ref{ss:lowres} in comparison with this previous work. Then, the relaxation of the lower limit demonstrates why the condensations of \citet{Kaneko:2015} were unable to collapse into a monolith but, instead, remained fairly diffuse. As explored in this work, the hydrostatic equilibrium within prominence condensations may have a significant role to play in the internal structuring and evolution of these condensations. Indeed, the pressure scale height (eq. \ref{eq:hdequilibrium}) for an atmosphere of hydrogen at 0.3~MK is $\approx$~9~Mm vs. 200~km for $T\approx6,000$~K - the temperature of the monolithic structure in Figures~\ref{fig:condensation_evol_low_res}~\&~\ref{fig:rti_evol_low_res}.

Figure~\ref{fig:flux_feeding} details the formation of subordinate flux ropes during the formation of the main 3G and 10G flux ropes. The exaggerated size of these structures has already been attributed to the large value of $\eta$ adopted within the higher-resolution cases. Nevertheless, the previous studies of \citet{Zhao:2017,Zhao:2019} simulated how chromospheric plasma can be levitated in the bottom part of a forming flux rope, which was formed in a similar process as studied here. Indeed, this is also further corroborated by the observational studies of \citet{Liu:2012b} and \citet{Zhang:2014,Zhang:2020}. In this work we find that the subordinate flux rope axes are relatively void of material, note the decrease in density in each case. Two relatively minor enhancements in density are, however, located at both the bow and in the wake of the rising ropes \citep[an observational feature commonly associated with flux ropes erupting from the solar atmosphere and forming \acp{CME}, cf. `three-part structure',][]{Dere:1997}. Such evolving \acp{CME} are commonly likened to snow plows in how they appear to scoop up material in the path of their eruption. The material that trails behind the core of the subordinate ropes is then akin to the prominence material in the core of an erupting or stable rope. Since it is much hotter material, it  would lack sufficient density or pressure to produce a characteristic absorption \citep[a feature rectified with the inclusion of a cool chromosphere as in ][]{Zhao:2017,Zhao:2019}.

Upon the condensation of prominence plasma at locations away from the topological dips, the localised density enhancements start sliding downwards, triggered by $N_{m, p}^{2}<0$ regions in each case. Figure~\ref{fig:visualisation} details the motion of those condensations that contain sufficient density and pressure (therein opacity) to produce an absorption or emission signature according to the \ac{halpha} visualisation method of \citet{Heinzel:2015}. In each case, the filament visualisations indicate significant oscillatory motions that are not a consequence of some global \ac{MHD} wave mode \textit{e.g.}, kink/sausage but instead simple oscillatory motions along a relatively-rigid magnetic field topology. The filament visualisations also indicate these condensations originate on either side of the flux rope axis and flow along flux surfaces irrespective of helicity. According to the prominence visualisations with O-point overlaid, in many cases these condensations can have initial heights above the central flux rope axis before ultimately collecting in the bottom of the flux rope. Furthermore, the motions of the smaller, more concentrated condensations in the latter stages of the 3G case are \textit{clearly} visible within the \ac{halpha} visualisations. This suggests these \ac{TI}-formed structures of dimensions $\approx~0.07''$ will potentially be observable using the unprecedented spatial and temporal observations available with the upcoming \ac{DKIST} instrument suite (diffraction-limited to 0.011$''$ per pixel).

Each of the prominence formation cases explored in this manuscript ends up in a different state from the others. The lower-resolution case explored in Section~\ref{ss:lowres} describes a monolithic structure that was slowly losing mass from the main body via isolated but repeated instances of reconnection. The 3G case does not display these small-scale islands falling from the underside but instead suggests the entire prominence itself may be falling through the host flux rope. In the wake of the falling prominence we see signatures of reconnection \textit{i.e.}, topology changes, in addition to localised heating that propagates throughout the core of the flux rope, heating it to temperatures in excess of 4~MK. The 10G case was not advanced in time to the same evolutionary state as the previous two but the final state of the 10G simulation suggested that the plasma was having a minimal effect on the topology of its host magnetic field. Consequently we do not see the core of the flux rope undergo sudden heating, instead the core is already at a temperature above 4~MK and the slight increase we do observe over time is most likely due to the dissipation of concentrated electric current as Joule heating. This hot, rather than cool, prominence cavity is consistent with the observations of hot X-ray coronal cavities in the literature, including  a recent statistical study by \citet{Bak-Steslicka:2019} \citep[see also \textit{e.g.},][although we note for the comparison with \citet{Fan:2006} that the simulations presented here do not contain the \ac{BPSS} configuration, but the lowest concave-up dips of the flux rope are indeed close to the bottom of the simulation domain]{Hudson:1999, Fan:2006, Reeves:2012}.

The aforementioned difference between the 3G and 10G cases is also well captured in Figure~\ref{fig:visualisation}. We see that for the 3G case, the O-point of the flux rope initially oscillates as a result of the formation process before the condensation of the prominence interrupts this bulk oscillation, as indicated with the clear displacement of the O-point to lower heights \citep[cf. Shafranov Shift,][]{Blokland:2011a, Jenkins:2019}. Strikingly, the position of O-point and prominence material in time appear completely correlated between $2800\lessapprox t\lessapprox4500$~s before the initiation of the internal reconnection that allows both the prominence material to continue falling and the O-point to return to a height of $\approx$~15~Mm \citep[an evolution that may prove related/comparable to eruptions that involve the ejection of a flux rope but leave the prominence behind \textit{e.g.},][]{Gilbert:2001, Jenkins:2018}. The 10G case, on the other hand, also displays oscillations related to the formation process but upon the appearance of condensations the O-point appears comparatively unshifted. Considering the masses of all cases, the lower resolution example attains a mass of $\approx$~2650~g/cm, in comparison with $\approx$~5000~\&~3500~g/cm for the 3G and 10G cases, respectively. Assuming a prominence with a characteristic length of 100~Mm such values would equate to total prominence masses of the order 10$^{13}$~g - an order of magnitude less than typically observed, two orders of magnitude less than some of the larger observational examples \citep[][]{Labrosse:2010, Vial:2015}. Hence, although the perturbation to the O-point is small in the 10G case, the prominence supplying this perturbation is correspondingly light.

The small masses of the prominences formed within this study are to be expected as it has long been known that the mass of a prominence cannot solely be supplied by the condensation of material already residing within a flux rope formed within the corona \citep[\textit{e.g.},][]{Saito:1973}. Instead, some additional ongoing mass-supply process is required for prominences of the order 10$^{14\,-\,15}$~g to form. \citet{Zhao:2017} presented a numerical study of the formation of a flux rope + prominence system wherein the cool chromospheric plasma was scooped up during the formation (reconnection) process (cf. the flux-feeding of Figure~\ref{fig:flux_feeding}). These authors report that their prominence attained a mass almost one order of magnitude higher than we find here. Furthermore, the study by \citet{Xia:2016} described the highly dynamic and ongoing formation and draining of plasma across a fully-3D flux-rope + prominence system. These authors, too, were only able to create a prominence of order 10$^{13}$~g. The studies by \cite{Fan:2017,Fan:2018,Fan:2020} have been able to produce prominences of order 10$^{15}$~g but the authors concede that this is due to the use of an enhanced heating term implemented in the photosphere that drives a significant amount of plasma into a very large ($>500$~Mm) flux rope.

And herein lies the particular problem. In order to obtain large, but observed, prominence masses, simulations are so far required to use empirical and unconstrained modifications to the physical evaporation processes to drive their formation \citep[][]{Fan:2017,Fan:2018,Fan:2020}. Adopting physical processes that are more readily constrained, the community find large masses but in the absence of particularly exotic dynamics - in particular with regards to the vertical motions observed within limb-projected prominences \citep[][]{Zhao:2017}. And finally, in the cases explored here and by \citet{Xia:2016}, the reproduction of the dynamic motions within prominences then limits the mass budget but also results in visualisations that are saturated in observations according to the methods of \citet{Heinzel:2015}. As such, it is clear that to address these inconsistencies, we must continue to pursue the understanding of the physical plasma and magnetic processes occuring within solar prominences, and do so with the combination of high-resolution and fully-3D simulations \citep[][]{Gibson:2018}.



\section{Summary of Conclusions}

Solar prominences often appear within the solar corona with little indication of \textit{how} the material got there. Multiple models have been proposed for this formation process, in this manuscript we have focussed on the spontaneous condensation model and characterised those processes responsible for the initial formation and subsequence evolution. Herein we have used a combination of advances provided by modern grid-adaptive simulations, to also describe evolutions at unprecedented resolutions. The \acf{CCI} has been shown to map well onto those flux surfaces that are slowly redistributing plasma in response to an evolving pressure gradient. Shortly thereafter, the \acf{TI} - driven by temperature decreases associated with optically-thin radiative losses - proves responsible for the initial formation of the cool condensations. Maps of the field-projected \acf{BV} frequency $N_{m, p}^{2}<0$ indicate these condensations form out of pressure balance with their surroundings and subsequently evolve towards the concave-up topological dips in the magnetic field. In the process of falling towards equilibrium, local misalignments between the pressure and density gradients within the condensations induce rotations as indicated with the maps of the baroclinicity metric. 
Therefore, we conclude that the initial formation and evolution processes for prominence condensations follows the ordering of \ac{CCI}\,--\,\ac{TI}\,--\,\ac{BV} \& baroclinicity. Upon the complete formation of prominence condensations, their evolution appears to be a consequence of the \ac{BV} and baroclinicity, with an apparent consistency with the mass-slippage theory of \citet{Low:2012a,Low:2012b,Low:2014}. In future work, we plan to revisit this scenario in full 3D, as already done at modest resolutions by \citet{Kaneko:2018}, where the extra dimension will suddenly allow for additional interchange and RTI routes to dynamical rearrangements. It will be of interest to then see how the condensations (mis)align with the overall helical field line structure within the flux rope, as this possibility to misalign density structures with the magnetic field was found to be unavoidable in idealized 2D and 3D local box simulations \citep{Claes:2019,Claes:2020}.

\begin{acknowledgements}
We wish to thank the anonymous referee for their constructive comments that improved the paper. We acknowledge the open source software that made possible the data visualisations presented within this work (\href{https://www.paraview.org}{Paraview};\href{https://www.python.org}{Python};\href{https://yt-project.org}{yt-project};\href{https://matplotlib.org}{matplotlib}). RK and JJ are supported by the ERC Advanced Grant PROMINENT and a joint FWO-NSFC grant G0E9619N. This project has received funding from the European Research Council (ERC) under the European Union's Horizon 2020 research and innovation programme (grant agreement No. 833251 PROMINENT ERC-ADG 2018). This research is further supported by Internal funds KU Leuven, project C14/19/089 TRACESpace. The computational resources and services used in this work were provided by the VSC (Flemish Supercomputer Center), funded by the Research Foundation Flanders (FWO) and the Flemish Government - department EWI.
\end{acknowledgements}


\bibliographystyle{aa} 
\bibliography{bibliography} 
\end{document}